\documentclass[twocolumn,floatfix,aps,superscriptaddress]{revtex4}
\usepackage{subfigure,dcolumn,jabbrv}
\usepackage{graphicx}
\usepackage{dcolumn}
\usepackage{bm}
\usepackage{CJK}
\usepackage{url}
\usepackage{color}
\usepackage{booktabs}
\usepackage{natbib}

\usepackage{amsmath,bm}
\usepackage[utf8]{inputenc}
\usepackage[british,UKenglish,USenglish,american]{babel}
\usepackage{bbm}
\usepackage{amssymb}
\usepackage{amsmath}
\usepackage{hyperref}
\usepackage{cleveref}
\usepackage{tikz}\usetikzlibrary{shapes.geometric, arrows}
\tikzstyle{startstop} = [rectangle, rounded corners, minimum width=3cm, minimum height=1cm,text centered, draw=black, fill=red!30]
\tikzstyle{io} = [trapezium, trapezium left angle=70, trapezium right angle=110, minimum width=3cm, minimum height=1cm, text centered, draw=black, fill=blue!30]
\tikzstyle{process} = [rectangle, minimum width=3cm, minimum height=1cm, text centered, text width = 4cm, draw=black, fill=orange!30]
\tikzstyle{decision} = [diamond, minimum width=3cm, minimum height=1cm, text centered, text width = 4cm, draw=black, fill=green!30]
\tikzstyle{arrow} = [thick,->,>=stealth]

\usepackage{graphicx}

\usepackage[normalem]{ulem}
\usepackage{appendix}
\usepackage{color}

\usepackage{pdfpages}
\usepackage{chngpage}

\def\den0{$\rho_0$}

\def\es0{$E_{\rm sym}(\rho_0)$}

\def\us0{$U_{\rm sym}(\rho_0,k_F)$~}

\def\l0{$L(\rho_0)$~}

\newcommand{\beq}{\begin{equation}}
\newcommand{\eeq}{\end{equation}}
\newcommand{\ba}{\begin{array}}
\newcommand{\ea}{\end{array}}
\newcommand{\bea}{\begin{eqnarray}}
\newcommand{\eea}{\end{eqnarray}}
\newcommand{\bi}{\begin{itemize}}  
\newcommand{\ei}{\end{itemize}}
\newcommand{\ben}{\begin{enumerate}} 
\newcommand{\een}{\end{enumerate}}
\newcommand{\bc}{\begin{center}}
\newcommand{\ec}{\end{center}}

\usepackage{pgf}

\usepackage{array}

\begin{document}

\title{Neural Network Emulation of Flow in Heavy-Ion Collisions at Intermediate Energies} 

\author{Nicholas Cox\footnote{ncox6@leomail.tamuc.edu}}
\affiliation{Department of Physics and Astronomy, Texas A$\&$M University-Commerce, Commerce, TX 75429, USA}
\author{Xavier Grundler\footnote{xgrundler@leomail.tamuc.edu}}
\affiliation{Department of Physics and Astronomy, Texas A$\&$M University-Commerce, Commerce, TX 75429, USA}
\author{Bao-An Li\footnote{Corresponding author: Bao-An.Li@tamuc.edu}}
\affiliation{Department of Physics and Astronomy, Texas A$\&$M University-Commerce, Commerce, TX 75429, USA}
\date{\today}

\begin{abstract}
Applications of new techniques in machine learning are speeding up progress in research in various fields. In this work, we construct and evaluate a deep neural network (DNN) to be used within a Bayesian statistical framework as a faster and more reliable alternative to the Gaussian Process (GP) emulator of an isospin-dependent Boltzmann-Uehling-Uhlenbeck (IBUU) transport model simulator of heavy-ion reactions at intermediate beam energies. We found strong evidence of DNN being able to emulate  the IBUU simulator's prediction on the strengths of protons' directed and elliptical flow very efficiently even with small training datasets and with accuracy about ten times higher than the GP. Limitations of our present work and future improvements are also discussed.
\end{abstract}

\maketitle

\section{Introduction}\label{intro}
Various neural networks (NNs) have been widely used as useful tools in machine learning to extract more connections within datasets and also as a way to generate or simulate physical events \cite{mcmc_ml}. Since the pioneering work of Ref. \cite{il-st}, promising results have been found in using NNs to address many critical issues in nuclear physics, see, e.g., Refs. \cite{Boe22,Zhou23,He23} for recent reviews. In the area of heavy-ion reactions at intermediate energies (around tens of MeV/nucleon to about 2 GeV/nucleon before the quark degree of freedom becomes important), interesting results have been found from applying NN techniques. For example, these techniques have been used in determining the centrality \cite{Bass-nn}, extracting nuclear symmetry energy \cite{WANG2022137508}, classifying orientation and deformation of colliding nuclei \cite{UU} on an event-by-event basis. One particularly useful feature of NNs is its potential applications in efficiently and reliably emulating solutions/outputs of computationally expensive and physically complicated models/processes, see, e.g., Refs\ \cite{Rap20,Pei21,Dean22,Love22,Mol22,Kno23,Nob23,Yang23,Skyrme_DL,Lay24} for recent examples in low-energy nuclear physics. A huge draw to NNs is the possibility that they need much shorter computation times than traditional methods \cite{Kasim_2022}. Indeed, very encouraging results and some challenges have been found in exploring such possibilities in nuclear physics, mostly in emulating nuclear structure and fission based on energy density functional theories so far \cite{Boe22,Zhou23,He23}. 

In this work, we explore the feasibility of using deep neural networks (DNNs) to emulate an isospin-dependent Boltzmann-Uehling-Uhlenbeck (IBUU) transport model simulator \cite{Li-Bauer,LCK} in predicting both the directed flow ($v_1$) and elliptical flow ($v_2$) \cite{pawel85,oll,art} in heavy-ion reactions at intermediate energies. Such an emulator is invaluable in further quantifying uncertainties of parameters characterizing the Equation of State (EOS) (e.g., incompressibility $K$) and transport properties (e.g., in-medium nucleon-nucleon cross section $\sigma^{med}_{NN}$) of hot and dense matter formed in heavy-ion reactions \cite{LRP,Hermann}. Within a Bayesian framework using a Gaussian Process (GP) \cite{GP} emulator trained with IBUU events generated by using 89 sets of input model parameters \cite{Li:2023ydi}, Li and Xie inferred the most probable incompressibility $K$ and the modification factor ($X$) of in-medium nucleon-nucleon cross sections compared to their free-space values from protons' directed flow (measured using the mid-rapidity slope $F_1$ of $v_1$) and elliptical flow in mid-central Au+Au collisions at $E_{\rm beam}/A =1.23$ GeV taken by the HADES Collaboration \cite{HADES1,HADES2}. 
While the GP emulator has been the standard emulation technique for model-to-data comparison in Bayesian analyses (Bayesian+GP) of heavy-ion collisions from low to ultra-relativistic beam energies, see, e.g., Refs. \cite{Scott1,Bass2,Scott2,Ber16,Scott3,Bass1,Weiss23,Kuttan,Hef23,wang2024,Phi21,JETSCAPE}, 
to reduce the interpolation uncertainty inherent in GP emulators, especially for large numbers of model parameters and multi-messenger observables, more efficient and accurate emulation techniques are being sought by the community. Moreover, generating a sufficiently large training dataset for GP can be computationally impractical. Therefore, alternative techniques to more efficiently and accurately emulate transport model simulators of heavy-ion reactions are certainly needed.

\begin{figure*}
    \centering
   \resizebox{0.495\textwidth}{!}{
  \includegraphics[]{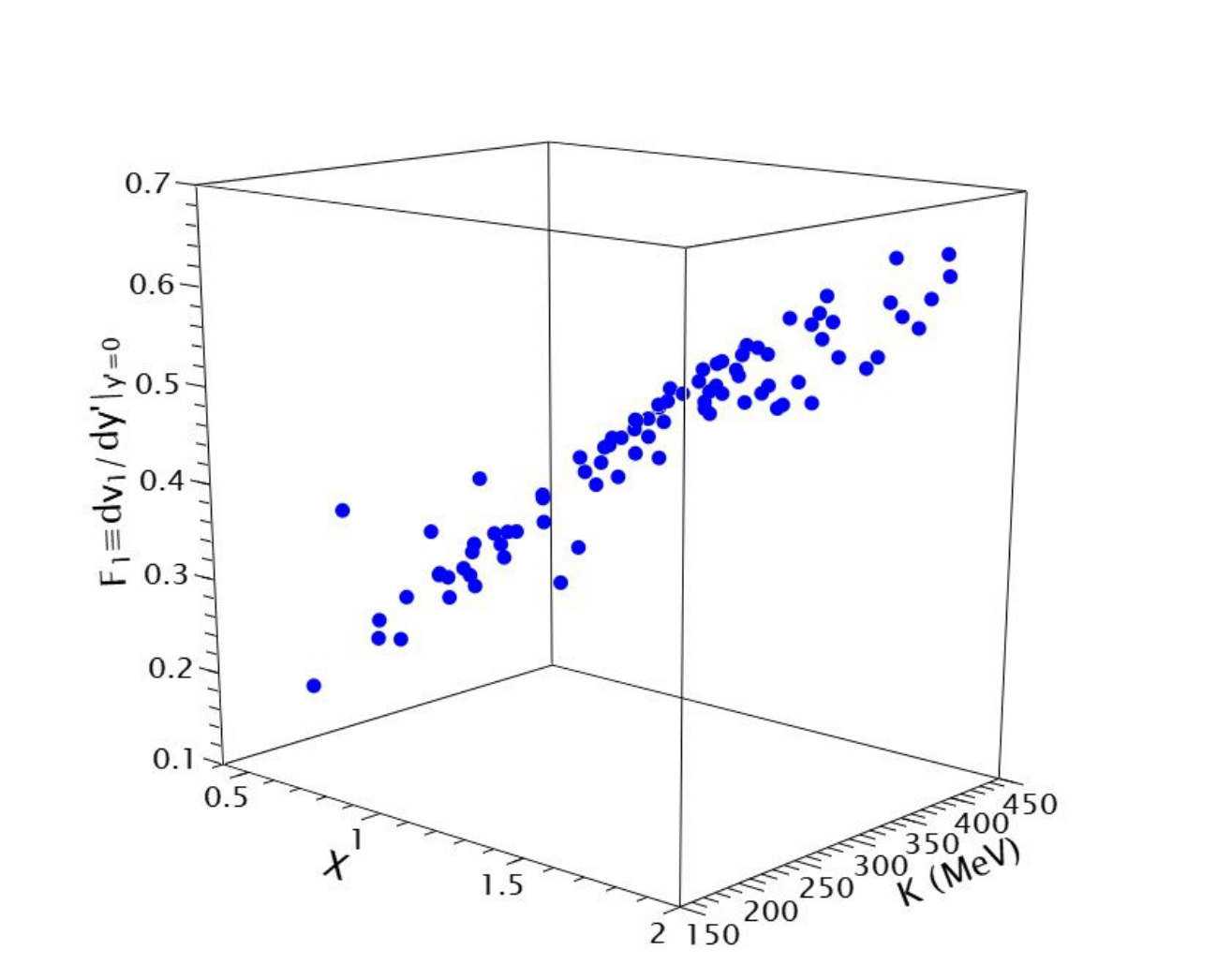}
  }
     \resizebox{0.495\textwidth}{!}{
   \includegraphics[]{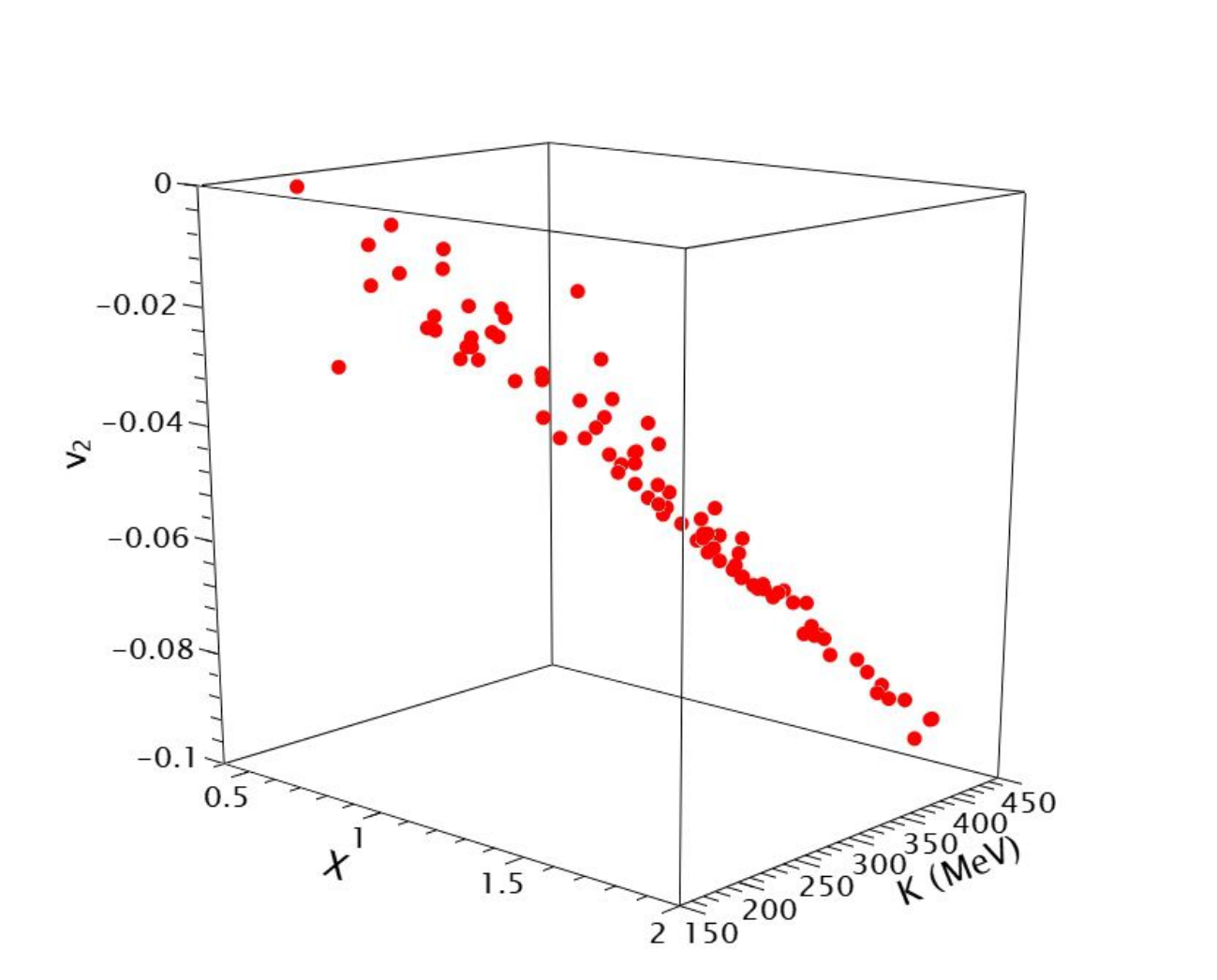}
   }
    \caption{89 sets of predicted slope $F_1$ of the free proton directed flow at mid-rapidity $dv_1/dy'|_{y'=0}$ (left) and elliptical flow $v_2$ (right), respectively, as functions of the incompressibility $K$ (right axis) and in-medium baryon-baryon scattering cross section modification factor $X$ (left axis) generated on a Latin Hyperlattice \cite{LH} in the X-K plane using the IBUU transport model for mid-central Au+Au reactions at 1.35 GeV/nucleon \cite{Li:2023ydi}.}\label{fig:f1v2scatter}
\end{figure*}

Here we focus on developing and evaluating a DNN as an alternative of GP in emulating heavy-ion reactions at intermediate beam energies. To determine if DNNs can replace all or part of the Bayesian+GP process with greater efficiency, we explore two possibilities. First, we test to see if DNN could replace the GP with greater efficiency, i.e. could we build a better emulator using the same amount of or less training events than GP? Second, we explore directly inferring the $X$ and $K$ values without the costly Markov Chain Monte Carlo (MCMC) process required for the Bayesian analysis. We found strong evidence of DNN being able to emulate very efficiently the IBUU simulator's results (predicting $F_1$ and $v_2$ given $X$ and $K$) even with small training datasets with an accuracy about ten times higher than the GP. 
On the other hand, the DNN's direct application in the inverse process is much less accurate. Therefore, a Bayesian+DNN inference will be a viable alternative to the more computationally costly Bayesian+GP approach.

The rest of the paper is organized as follows. In Section \ref{mm}, we explain the models we used and our methods for scoring them. We then present and discuss our results in Section \ref{res}. Finally, we summarize in Section \ref{sum}.

\section{Models and Methods}\label{mm}
\subsection{Inputs and outputs of the IBUU simulator}
For completeness and ease of our presentation, we briefly recall here the Bayesian+IBUU (GP) framework as well as the major inputs and outputs of the IBUU simulator used in Ref. \cite{Li:2023ydi}. 
Transport model simulations of heavy-ion reactions are computationally expensive and thus can't be used directly in generating normally multi-million steps in the MCMC sampling of posterior probability distribution functions (PDFs) of model parameters in Bayesian analyses.
Therefore, a GP emulator was used successfully in Ref. \cite{Li:2023ydi} to mimic predictions of the IBUU simulator in its simplest setup. As reviewed in detail in Refs. \cite{Li-Bauer,LCK}, the IBUU simulator normally has multiple choices for the baryon mean-field potential including its isospin and momentum dependence \cite{LiBA04A,LiBA04B} as well as in-medium baryon-baryon cross sections \cite{Li-Chen}. In its first application within the Bayesian+IBUU (GP) framework, only the two most important input parameters ($X$ and $K$) are randomly initialized in their prior uncertainty ranges while others are fixed at their currently known most probable values in each MCMC step, and no momentum-dependence of single nucleon potentials was considered \cite{Li:2023ydi}. The latter requires much more computing time but is very important for understanding several interesting features of heavy-ion collisions at intermediate energies. Once a faster and more reliable emulator is developed, this certainly should be considered. The two most important input parameters $X$ and $K$ for the IBUU simulator are defined as
\begin{enumerate}
    \item The in-medium baryon-baryon scattering modification factor 
$X\equiv\frac{\sigma^{med}_{NN}}{\sigma^{free}_{NN}}$ where $\sigma^{free}_{NN}$ is the free-space cross section. 
\item The incompressibility quantifying the stiffness of SNM EOS
$E_0(\rho)$ via $K=9\rho^2_0\left(\frac{d^2E_0(\rho)}{d\rho^2}\right)$ at its saturation density \den0$=0.16$ fm$^{-3}$. 
\end{enumerate}

The goal of Ref. \cite{Li:2023ydi} was to infer the PDFs of $X$ and $K$ from the experimental observations of nuclear collective flow. The observables chosen were proton directed (or transverse) flow ($v_1$) and elliptical flow ($v_2$). They are coefficients of the Fourier decomposition of particle azimuthal angle $\phi$ distribution $\frac{2\pi}{N}\frac{dN}{d\phi} = 1 + 2\sum_{n=1}^{\infty}v_n\cos{[n(\phi)]}$ where $\phi$ is measured with respect to the reaction plane in $x-o-z$ for a beam in the $z$-direction. Their values at rapidity $y$ and transverse momentum $p_t$ can be evaluated from
$v_1(y,p_t)=\left<cos(\phi)\right>(y,p_t)=\frac{1}{n}\sum_{i=1}^{n}\frac{p_{ix}}{p_{it}}$ and
$v_2(y,p_{t})=\left<cos(2\phi)\right>(y,p_t)=\frac{1}{n}\sum_{i=1}^{n}\frac{p_{ix}^2-p_{iy}^2}{p_{it}^2}$, where $p_{ix}$ and $p_{iy}$ are the x- and y-component of the $i^{{\rm th}}$ particle momentum, respectively. In Ref. \cite{Li:2023ydi} and in the community generally, rather than use $v_1$ directly, its slope $F_1 = dv_1/dy'|_{y'=0}$ at mid-rapidity in the center of mass (cm) frame is often used to measure the strength of directed flow. We will only use $F_1$ from here on. In mid-central (10-30\% centrality) Au+Au collisions at $E_{\rm beam}/A$=1.23 GeV measured by the HADES Collaboration \cite{HADES1,HADES2}, 
$F_1\equiv dv_1/dy'|_{y'=0}=0.46\pm 0.03$ and $v_2=-0.06\pm 0.01$ data for $|y_{cm}|\leq 0.05$ and $p_t\geq 0.3$ GeV/c were observed for free protons. These values are at approximately the maximum and minimum of the excitation functions of $F_1$ and $v_2$  accumulated over the last 40 years \cite{LRP}, respectively. Namely, the strengths of both directed flow (positive) and elliptic flow (negative) are the strongest around this beam energy. 
The $F_1$ and $v_2$ data from the HADES experiments are thus particularly interesting and informative. Indeed, these two experimental observations have already provided some strong constraints on the PDFs of $X$ and $K$ \cite{Li:2023ydi}.

In generating the IBUU training and testing datasets, for each set of the X and K parameters generated on a Latin Hyperlattice \cite{LH}, 200 testparticles/nucleon and 100 impact parameters $b$ with its probability density given by $P(b)\propto b$ between b=6 fm and 9 fm corresponding to approximately the (10-30)\% centrality are used. The accumulated 20,000 effective nucleus-nucleus collision events are used to evaluate the $F_1$ and $v_2$ for each $X-K$ parameter set. Shown in Figure\ \ref{fig:f1v2scatter} are the $F_1$ and $v_2$ values for 89 $X-K$ parameter sets used for training the GP emulator in Ref. \cite{Li:2023ydi}. This dataset will also be used in this work.

\subsection{A deep neural network as an emulator of IBUU}
A neural network is like a regression, except that it uses perceptrons. A perceptron is an individual neuron or point, that can have values varying between 0 and 1. The output is based on the activation function of the perceptron as well as its weights and biases. If the perceptron reaches a certain value or higher, based on its inputs, then it will activate. The neural network will automatically adjust the weights and biases of each perceptron during the training phase to converge on some set of values that can solve given inputs for their corresponding outputs. Neural networks have layers of perceptrons, starting with the input layer, where the data goes in, then hidden layers, which is where the computation happens and the user has little control while the machine calculates the best possible values of weights and biases, and ending up with the output layer. The goal of the neural network in training is to minimize the loss between its outputs and the actual outputs using a loss function.

We used two different, popular Python packages to create DNNs in order to determine which would be easiest and most efficient. The first is Scikit-Learn \cite{scikit-learn,sklearn_api}. The second is TensorFlow \cite{tensorflow2015-whitepaper}, which runs on the Keras API \cite{chollet2015keras}. It should be noted that TensorFlow is designed to be run on GPUs, but can be used with CPUs as was done in this study.

\begin{figure}
    \centering
    \includegraphics[width=0.75\linewidth]{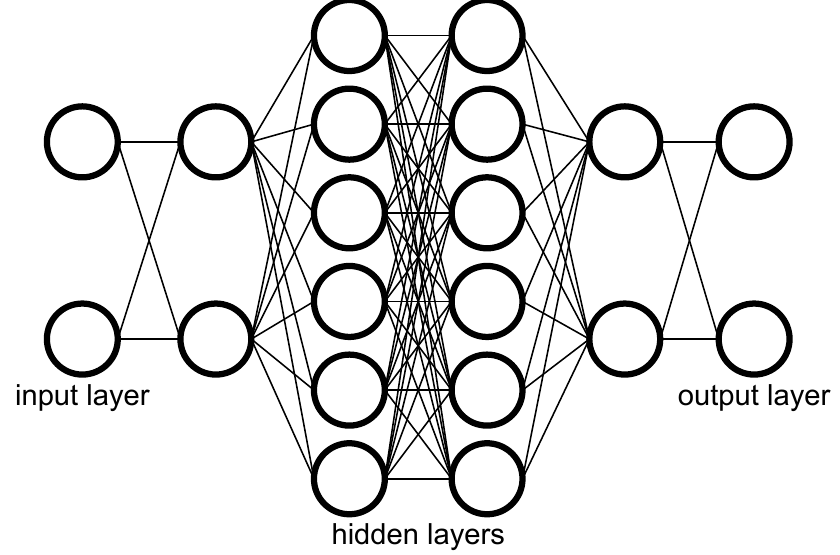}
    \caption{Diagram of DNN with four hidden layers}
    \label{fig:nnDiagram}
\end{figure}

All models used four hidden layers as shown in Figure \ref{fig:nnDiagram}. The number of perceptrons in the first and last hidden layers were chosen to match the number of inputs and outputs. The two middle layers each had six perceptrons, which allowed the DNN to vary more weights and biases in order to better optimize results. All layers are interconnected, i.e. dense. This means the output of each perceptron is passed to each perceptron in the next layer. Figure \ref{fig:nnDiagram} shows this graphically. We set most layers to use the hyperbolic tangent function (tanh) as their activation function except the first and last hidden layers in the TensorFlow models, which allowed for more variability in the layers. For these, we used a linear activation function. Except where mentioned, all data was scaled using Scikit-Learn's StandardScaler (SS), which scales each feature (i.e. $F_1$ and $v_2$) separately to have a zero mean and unit variance. The scaled data were used for training the DNN and calculating $R^2$, for all other purposes we report the unscaled values. The Scikit-Learn models used the multi-layered perceptron regressor (MLPR) with the Limited-memory Broyden-Fletcher-Goldfarb-Shanno optimizer (L-BFGS) \cite{Liu:1989esw}, which, according to the Scikit-Learn User Guide, is well-designed for small datasets. The TensorFlow models used a Keras Sequential model with the Adam optimizer \cite{kingma2017adammethodstochasticoptimization}. The loss function across all models was the mean squared error (MSE) between the DNN output $X_{\rm{obs,i}}$ and the expected/experimental/actual output $X_{\rm{exp,i}}$:
\begin{equation}
    {\rm MSE} = \frac{1}{n} \sum^{n}_{i = 1} (X_{\rm{obs,i}} - X_{\rm{exp,i}})^2
\end{equation}
where n is the number of observations made. For the Scikit-Learn models, we manually adjusted the max number of iterations to optimize the DNN predictions as measured by $R^2$:
\begin{equation}
    R^2=1-\frac{\rm{SSE}}{\rm{TSS}}
\end{equation}
where SSE is the sum of squared errors (also sometimes written as RSS for the residual sum of squares) and TSS is the total sum of squares as discussed in Ref. \cite{Richter:2023zec}. For the TensorFlow models, we manually adjusted the number of training epochs for the same goal. All other model parameters were left at their default values. Data was split 75\% for training and 25\% for validation.

\begin{figure*}
    \centering
    \begin{tabular}{cc}
        \includegraphics[width=0.4\linewidth]{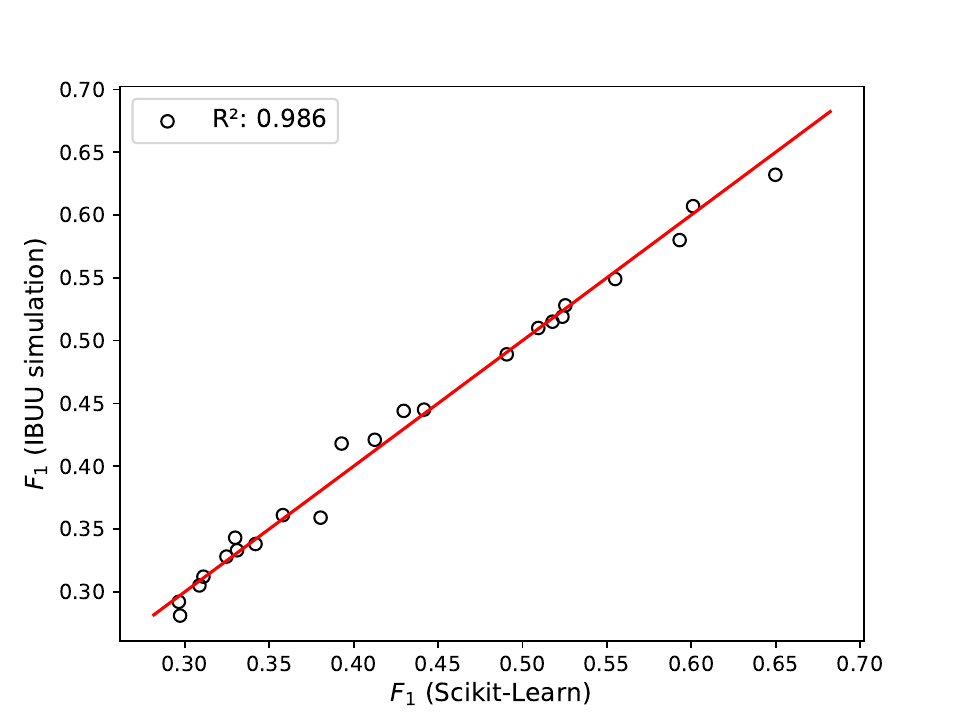} &
        \includegraphics[width=0.4\linewidth]{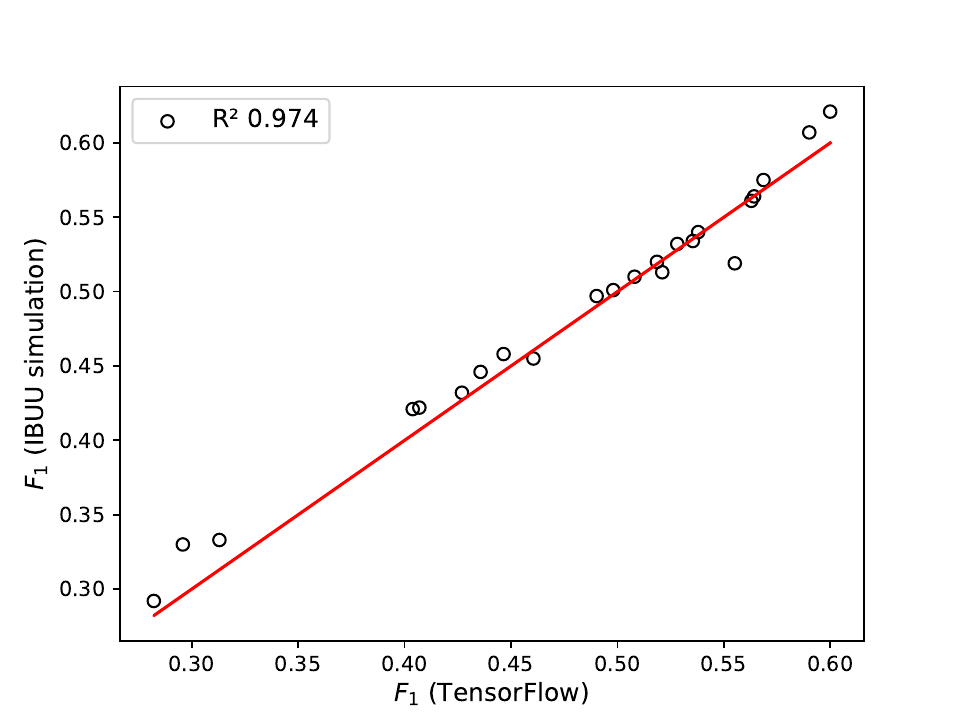} \\
        \includegraphics[width=0.4\linewidth]{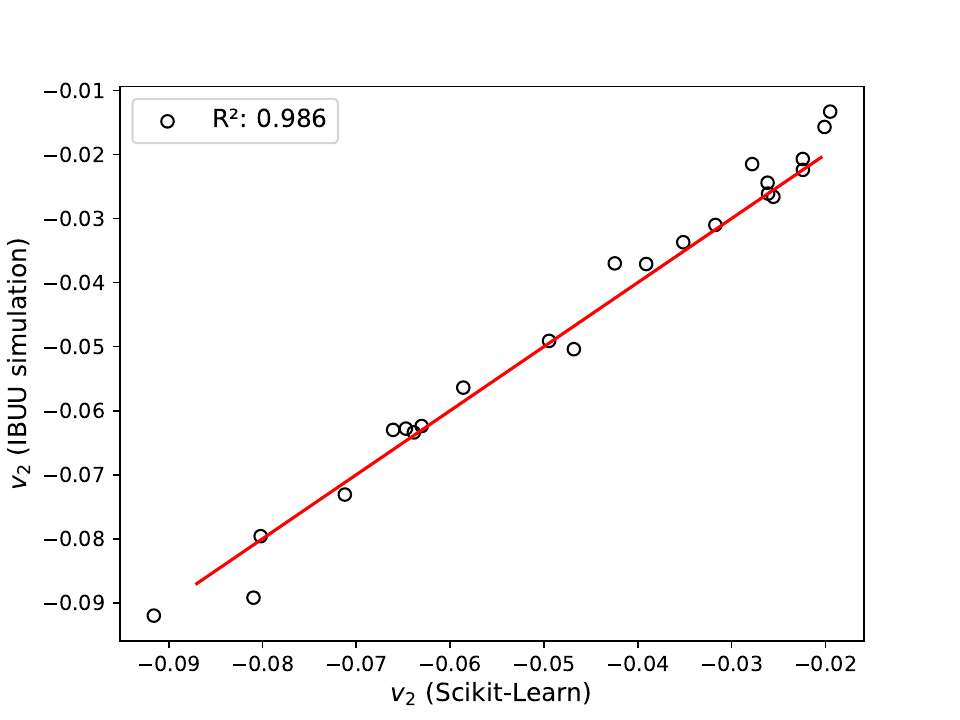} &
        \includegraphics[width=0.4\linewidth]{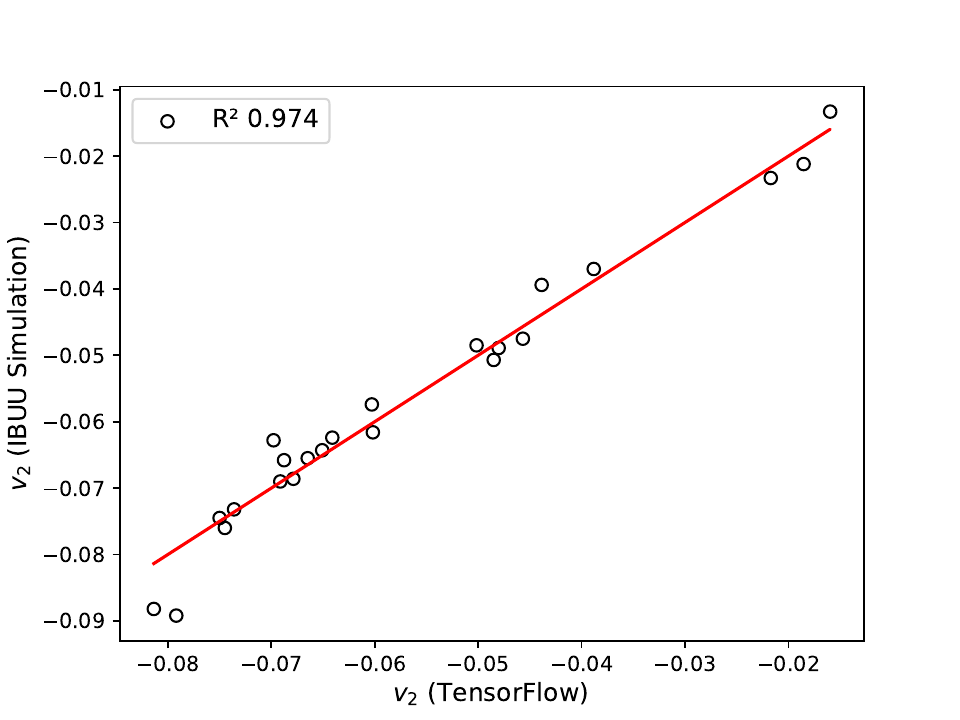} 
    \end{tabular}
    \caption{DNN predictions are black circles. The line of the perfect match between the DNN predictions and the IBUU simulations is red. }
    \label{fig:v1v2Pred}
\end{figure*}
To gauge the success of our DNN, we first got some analytical results to be used as references. As discussed in Ref. \cite{Li:2023ydi} and shown in Figure \ref{fig:f1v2scatter}, $F_1$ and $v_2$ have an approximately linear relationship with respect to X and K. As a check on this, we performed a multiple regression assuming a linear fit on the simulation data. The high $R^2$ corroborates the previous observation. Thus, it is expected that predicting $F_1$ or $v_2$ from X and K should not be difficult. To see if inferring the model parameters from the observables would be as trivial, the same linear multiple regression was done with $F_1$ and $v_2$ as the independent variables and X and K as the dependent variables. Not surprisingly, simply reversing the problem did not yield excellent results in terms of the $R^2$ values. Table \ref{tab:regressTab} summarizes the results for all regressions. These will be used as the minimum success level for the DNN in our work. Ideally, the models will perform better (have a higher $R^2$ value) than these basic linear approximations.

\begin{table}[ht]
    \centering
    \begin{tabular}{c|c}
        Equation & $R^2$\\\hline
        $F_1 = -0.0251 + 0.206\cdot X + 0.000663\cdot K$ & 0.927\\
        $v_2 = 0.0519 - 0.0400\cdot X - 0.000158\cdot K$ & 0.959\\
        $X = -0.130 + 2.92\cdot F_1 - 1.34\cdot v_2$ & 0.782\\
        $K = 403 - 881\cdot F_1 - 6400\cdot v_2$ & 0.587\\
        \hline
    \end{tabular}
    \caption{Results of regression based on IBUU predictions}
    \label{tab:regressTab}
\end{table}

\begin{figure*}
    \centering
    \begin{tabular}{cc}
        \includegraphics[width=0.5\linewidth]{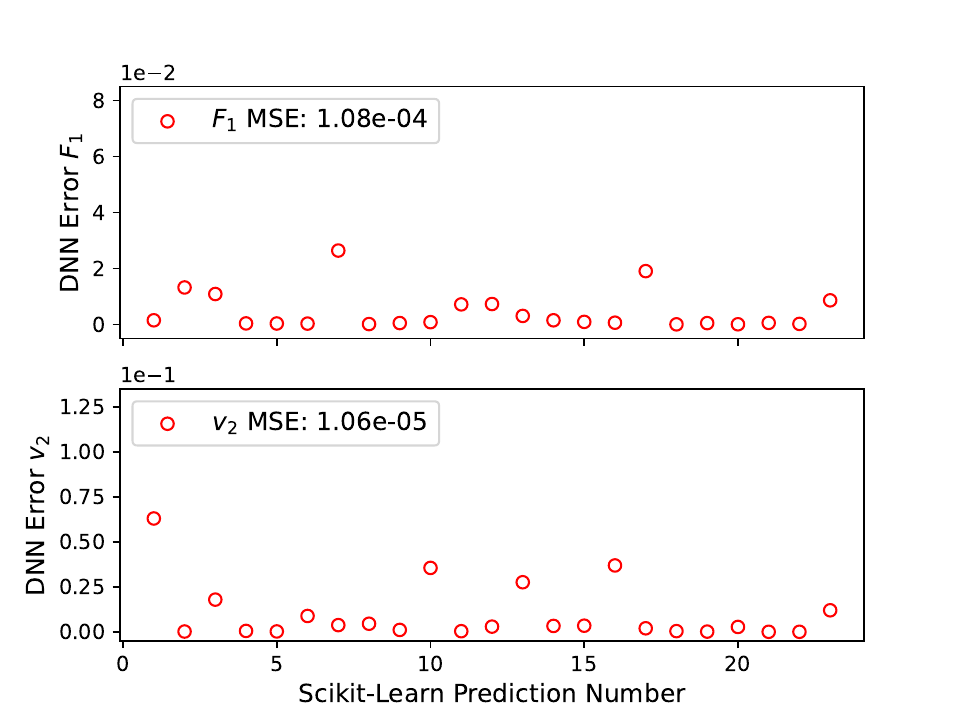} &
        \includegraphics[width=0.5\linewidth]{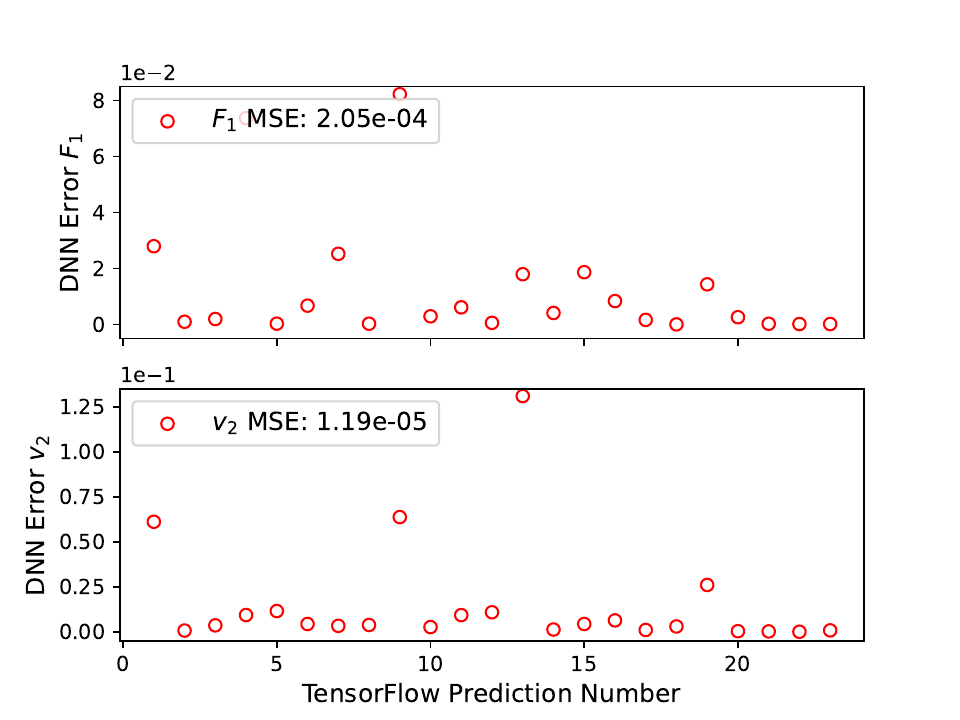}
    \end{tabular}
    \caption{DNN error for predicting $F_1$ and $v_2$, respectively, for each point in the test dataset shown in Fig. \ref{fig:v1v2Pred}.}
    \label{fig:err1}
\end{figure*}
\section{Results and Discussions }\label{res}
During our attempts to optimize the DNN structure, we noticed a great deal of inconsistency in our models' training. Without changing the code or DNN structure, the $R^2$ value could change significantly, as much as 0.4 with certain setups. Possible reasons for this are discussed later. Below we report our best results with quantified variations. Once we decided upon a DNN structure that produced high $R^2$ values most often, we ran the same code five times and picked the best. To quantify the consistency of each model, we report the range $\Delta$R$^2$ of the DNN predictions compared to the IBUU simulations for the test dataset in Table \ref{tab:consistency}.

\begin{table}[ht]
    \centering
    \begin{tabular}{c|c|c|c|c}
        Model & $\Delta R^2$ & Train Time(s) & Prediction Time(s)\\\hline
        SL emu & 0.109 & $5.44\times10^{-2}$ & $3.86\times10^{-6}$\\
        TF emu & 0.226  &3.87 & $2.29\times10^{-3}$\\    
        SL inv & 0.126 &$7.23\times10^{-2}$ & $3.96\times10^{-6}$\\
        TF inv & 0.312  & 5.79 & $2.24\times10^{-3}$\\
        \hline
    \end{tabular}
    \caption{SL is for Scikit-Learn, TF is for TensorFlow, emu (emulation) is for the DNN models emulating the IBUU simulation, and inv (inversion) is for the DNN models that inferred X and K from $F_1$ and $v_2$.}
    \label{tab:consistency}
\end{table}

Also included in Table \ref{tab:consistency} is the average time to train the DNN model and the average time per prediction on a PC. While no training is prohibitively expensive to do once, prediction time is important. With over a million predictions, the difference between Scikit-Learn and TensorFlow is over half an hour, which is important to consider in an emulator. As a reference, we notice that Ref. \cite{Li:2023ydi} used 50 million steps (calls to the GP emulator) in their MCMC process. This would require a total of less than four minutes of computation time on a PC by the Scikit-Learn DNN. Thus, DNN can be computationally efficient enough to replace the GP.

\begin{figure*}
    \centering
    \begin{tabular}{cc}
        \includegraphics[width=0.4\linewidth]{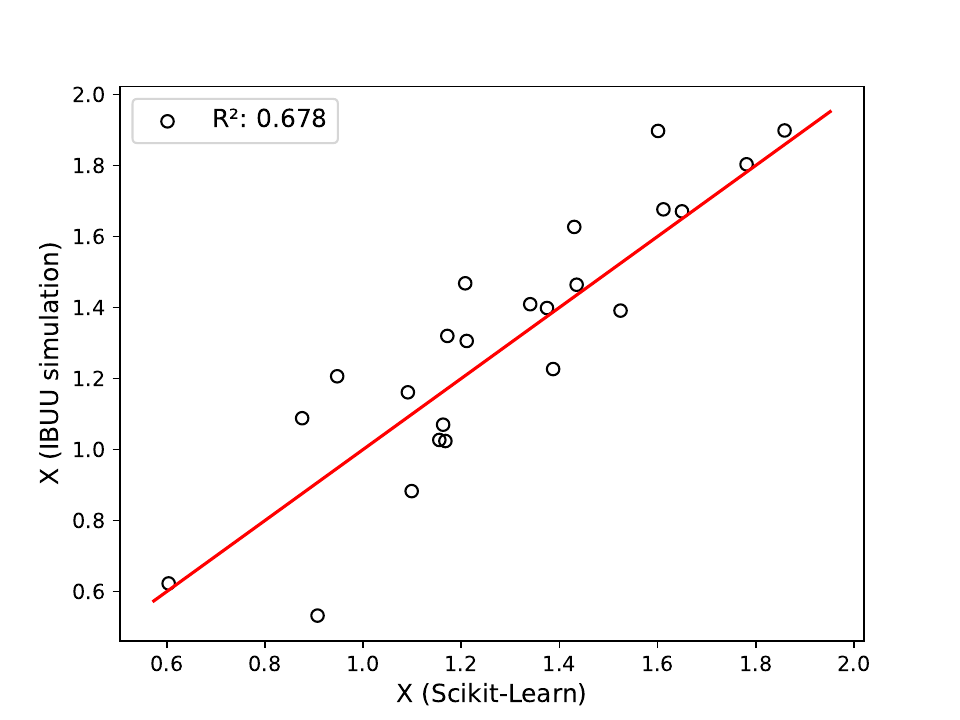} &
        \includegraphics[width=0.4\linewidth]{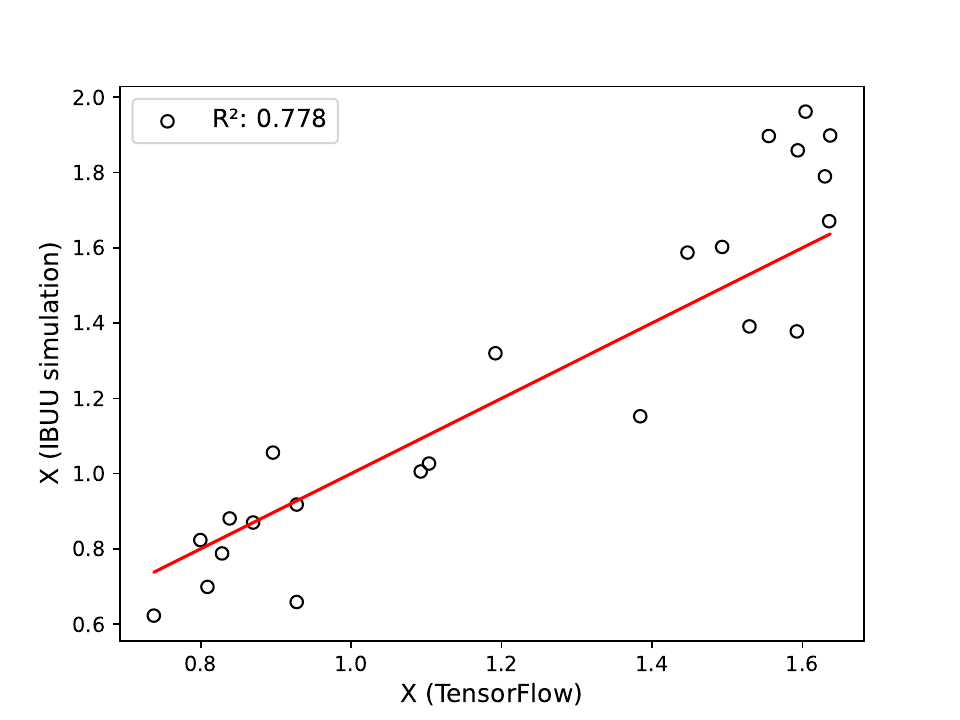} \\
        \includegraphics[width=0.4\linewidth]{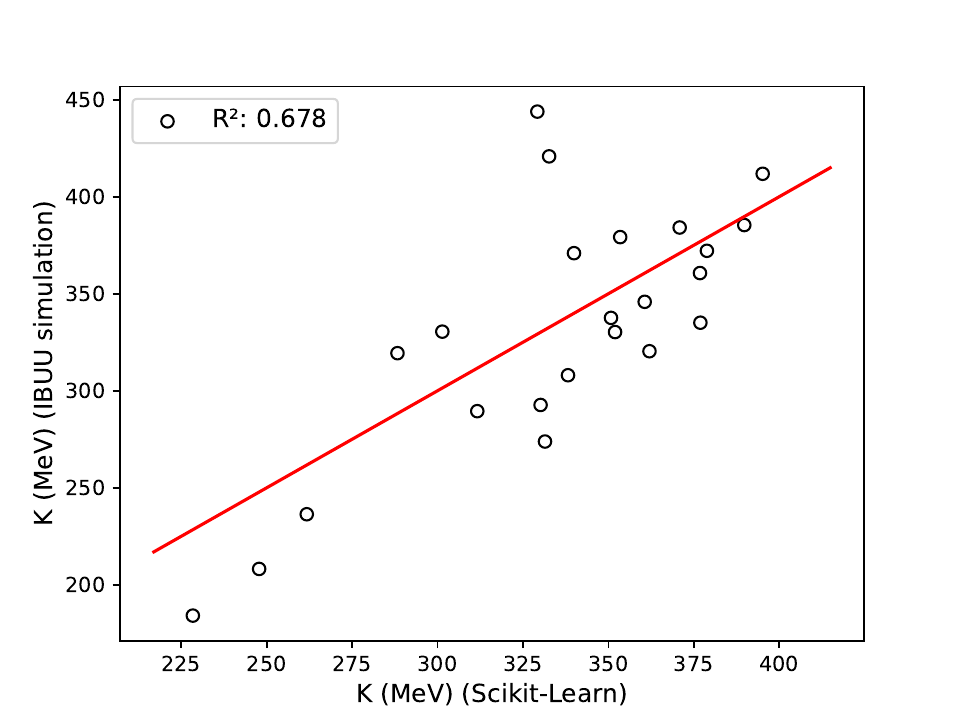} &
        \includegraphics[width=0.4\linewidth]{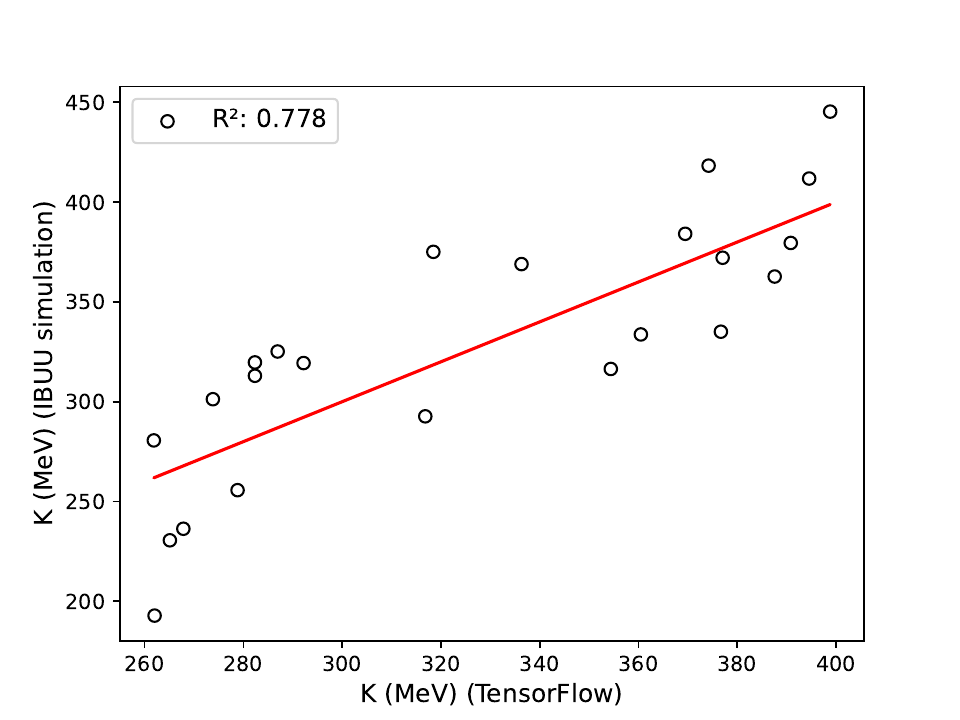} 
    \end{tabular}
    \caption{Same as Figure \ref{fig:v1v2Pred}, but now for X and K (MeV).}
    \label{fig:xkPred}
\end{figure*}

\begin{figure*}
    \centering
    \begin{tabular}{cc}
        \includegraphics[width=0.5\linewidth]{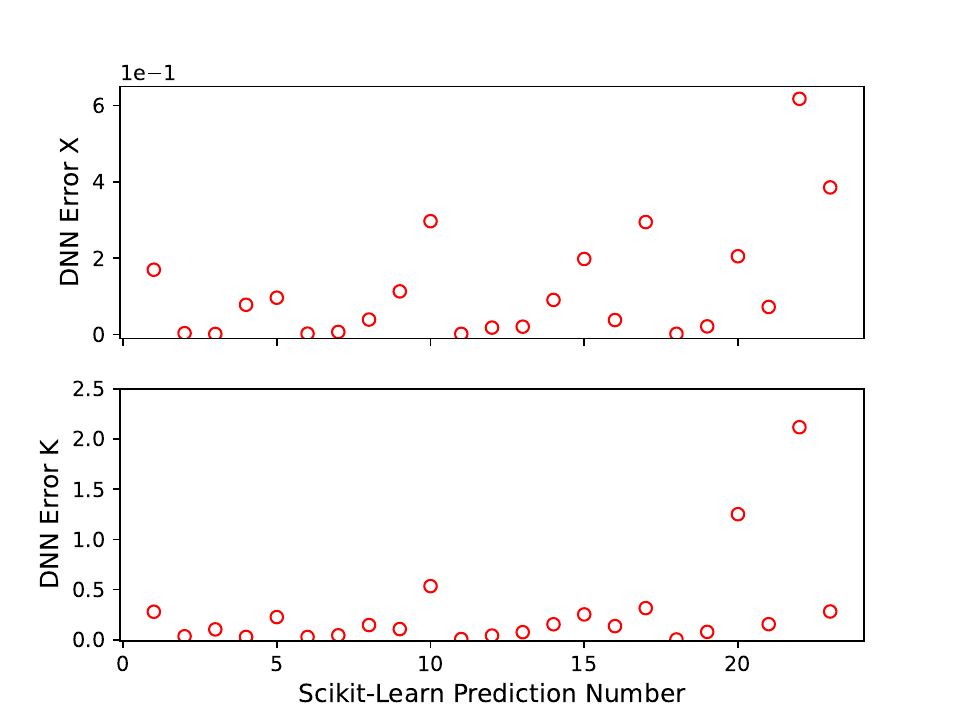} &
        \includegraphics[width=0.5\linewidth]{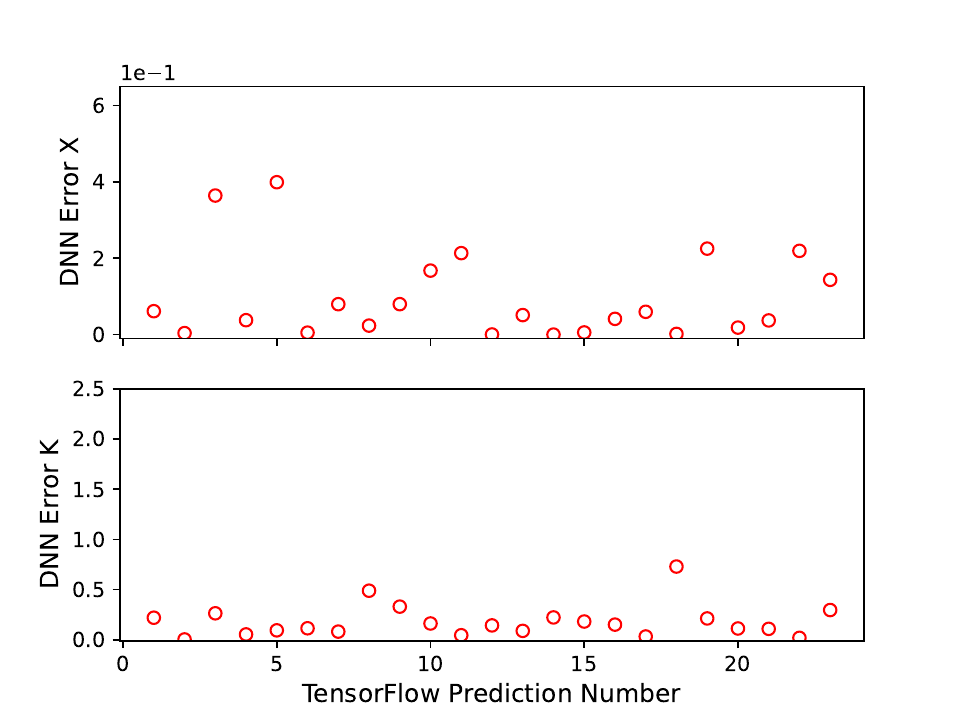}
    \end{tabular}
    \caption{DNN error for predicting X and K, respectively, for each point shown in Fig. \ref{fig:xkPred}.}
    \label{err2}
\end{figure*}
\subsection{Emulation}\label{emu}
Figure \ref{fig:v1v2Pred} summarizes the results from the DNN for predicting $F_1$ and $v_2$ from X and K. For the Scikit-Learn model, instead of using SS, a different scaler was used: the Max Absolute Scaler (MAS), which, according to the Scikit-Learn User Guide, is better for sparse data because it does not shift or center the data. Rather, it divides each feature by its maximum absolute value. We did not use it regularly, however, because in most cases it failed to improve performance and caused the TensorFlow models to become even more inconsistent. For these results, the max number of iterations was set to 80, an unusually low value. This prevented the model from converging every time, but allowing more time often caused over-fitting. The TensorFlow model was given 150 epochs to train. The high $R^2$ values show that both the Scikit-Learn and TensorFlow models improved on the linear regression.

To quantify the errors of the emulators, we define here the DNN Error for an arbitrary variable X as
\begin{equation}
    {\rm DNN~Error}_{{\rm run}}(x)=\frac{[X({\rm DNN})_{{\rm run}}-X({\rm IBUU})_{\rm run}]^2}{\sigma^2_X(\rm DNN)+\sigma^2_X(\rm IBUU)}
\end{equation}
where $\sigma^2_X(\rm DNN)$ and $\sigma^2_X(\rm IBUU)$ are the variances of X from the DNN emulator and IBUU simulator, respectively.
In essence, it measures the individual MSE of each variable X with respect to the total error of simulations and emulations involved, enabling a more fair comparison of the accuracies obtained with different approaches. In Figure \ref{fig:err1}, we report the DNN Error of the observable $F_1$ and $v_2$, respectively, for each point shown in Fig. \ref{fig:v1v2Pred}. 
Comparing the DNN Errors obtained with the Scikit-Learn and TensorFlow, the latter generally has larger errors in predicting both $F_1$ and $v_2$. More quantitatively, the largest DNN error for $F_1$ ($v_2$), indicating the point the DNN has the weakest predictive power, is $2.64 \times 10^{-2}$ ($6.30 \times 10^{-2}$) for Scikit-Learn versus $8.22 \times 10^{-2}$ ($1.31 \times 10^{-1}$) for TensorFlow. The absolute MSE values in predicting $F_1$ and $v_2$ are also indicated in Fig. \ref{fig:err1}. The results show strong signs of DNN being able to accurately emulate the IBUU simulator's predictions even with the small training dataset. Moreover, by comparing the DNN's MSE values with that of GP (Fig.\ 3 of Ref. \cite{Li:2023ydi}) trained by using the same set of IBUU predictions, we notice that overall the DNN is about ten times more accurate. More quantitatively, as shown in Fig. \ref{fig:err1}, the DNN total MSE is $1.19\times 10^{-4}$ with Scikit-Learn and $2.17\times 10^{-4}$ with TensorFlow, respectively. While the GP total MSE shown in Fig.\ 3 of Ref. \cite{Li:2023ydi} is $9.20 \times 10^{-4}$. 

\subsection{Inversion}\label{inv}
Unsurprisingly, the direct inversion of the problem did not produce such stunning results, and, unfortunately, the models did not make significant improvement over the simple linear multiple regression. Figure \ref{fig:xkPred} summarizes the results of trying to infer X and K from $F_1$ and $v_2$. Shown in Fig. \ref{err2} are the corresponding DNN errors in predicting X and K, respectively, for each point shown in Fig. \ref{fig:xkPred}. While the Scikit-Learn models were again stopped short of convergence at 150 iterations, the TensorFlow models actually did best at a slightly longer 250 epochs. It can be seen qualitatively that the models did a worse job predicting K than they did X. The $R^2$ values are the same because they score the entire model, not X and K predictions separately. That result is not surprising, however, given that the values for $R^2$ from the multiple regression were much smaller for K than for X. It is also seen that the Scikit-Learn and TensorFlow models have roughly the errors for inferring both X and K.

In the inversion problems, all models had poor accuracy, so concrete conclusions are difficult. Two possibilities for the issue seem likely. One, $F_1$ and $v_2$ simply do not carry enough information to constrain X and K tightly. It is well known that a degeneracy (i.e., combinations of small/big K with big/small X have the same physical effects) exists between these two parameters in producing nuclear collective flow, see, e.g., Refs. \cite{Ber88,Zheng99,LiSustich,Dan02,BALI2,Herman1,Zhang07,PLi18,Li22,Gale,Fuchs01}, which complicates inference problems. This degeneracy arises because the nuclear collective flow is generated by the pressure gradient during nucleus-nucleus collisions. And the pressure gradient can be created by either a relatively long-range nuclear force through a mean-field potential (its stiffness is characterized by K) or a short-range nuclear force through nucleon-nucleon scattering (its frequency and effect are characterized by X). Perhaps some other observable could help. This explanation would also account for the low $R^2$ values from the multiple regressions. The second reason is a lack of training data. While it appears to be possible to train DNN for emulation on only 89 data points, it is possible that more data is required for the more complex inversion problem. An idea for getting around the lack of IBUU simulation data is to first train a DNN as an emulator to produce large amounts of data, and then use this to train the inversion network. One danger in this is that any small error at the beginning of the process could be propagated and compounded through several models. Future work could assess the feasibility of this and quantitatively evaluate the systematic error. The above experience gained in using directly the DNN as an inversion tool is very educational. Nevertheless, we emphasize that the value of DNN designed here is not for direct inversions but as an emulator for solving inverse problems within the Bayesian+DNN framework.

\subsection{Effects of Training/Testing Data Splitting}\label{split}
Up to this point, we have used 75\% of the 89 data points for training all of our models and 25\% to test their accuracy. This 75/25 split, or similar, such as 80/20 or 70/30, is commonly used for designing DNN, but it is not a hard and fast rule, see, e.g., Ref. \cite{Joseph_2022} and references therein.

To determine if a different splitting ratio could improve our results, we tried two other scenarios: 50/50 and 90/10 training/testing ratio. In both cases, consistency remained elusive, so each was run five times. We report the maximum $R^2$ value each model achieved in Table \ref{tab:splitR2}.
\begin{table}[ht]
    \centering
    \begin{tabular}{c|c|c}
        $R^2_{max}$ & 50/50 & 90/10 \\\hline
        SL emu & 0.958 & 0.969 \\
        TF emu & 0.930 & 0.968 \\
        SL inv & 0.677 & 0.656 \\
        TF inv & 0.655 & 0.786 \\
        \hline
    \end{tabular}
    \caption{Maximum $R^2$ value achieved in five runs of each DNN setup with training/testing splits of 50/50 and 90/10.}
    \label{tab:splitR2}
\end{table}

\begin{figure*}
    \centering
    \begin{tabular}{cc}
        \includegraphics[width=0.5\linewidth]{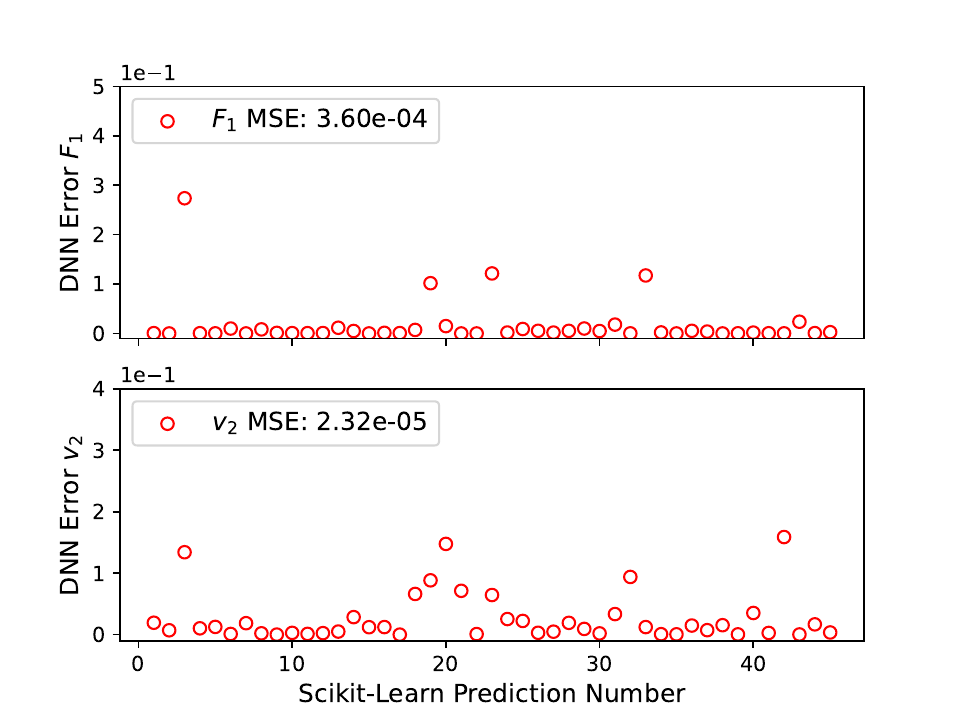} & 
\includegraphics[width=0.5\linewidth]{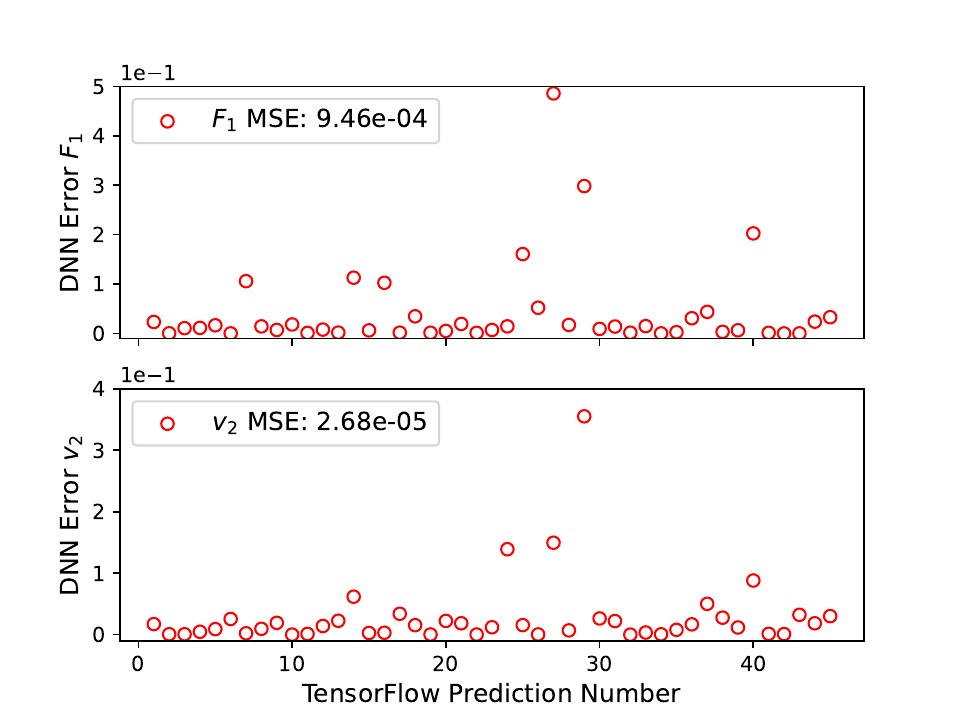}
    \end{tabular}
    \begin{tabular}{cc}
        \includegraphics[width=0.5\linewidth]{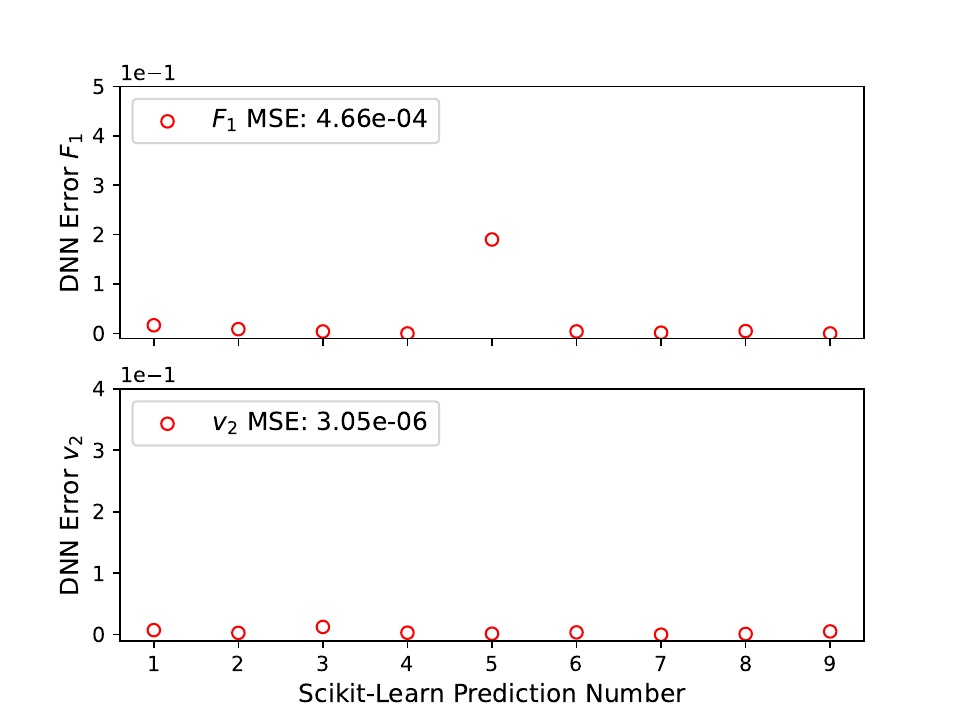} & 
\includegraphics[width=0.5\linewidth]{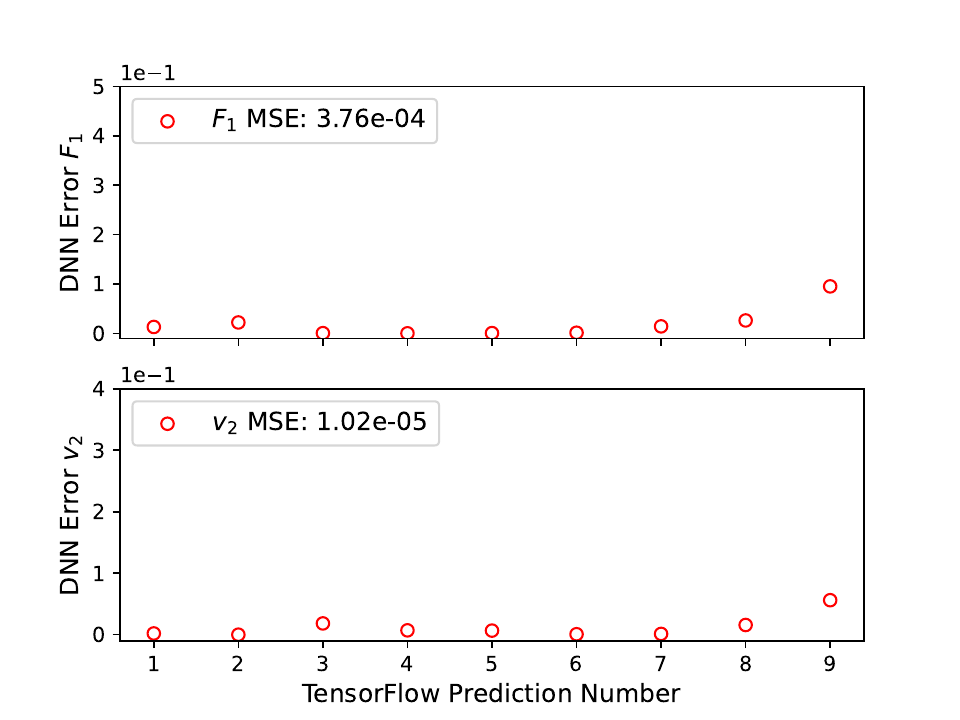} 
    \end{tabular}
    \caption{DNN error for predicting $F_1$ and $v_2$ with a 50/50 split (top) and 90/10 split (bottom).}
    \label{errSplit}
\end{figure*}

All attempts to infer X and K with the DNN continued to produce poor results, so we will not discuss them further.
Beginning with the 50\% training, 50\% testing case, we note that neither Scikit-Learn nor TensorFlow achieved $R^2$ values very close to our previous results, and this is reflected in the absolute MSE values shown in Fig. \ref{errSplit}. The total MSE for Scikit-Learn is $3.83 \times 10^{-4}$, more than three times worse than before, and $9.73 \times 10^{-4}$ for TensorFlow, which is now on the order of the GP used in Ref. \cite{Li:2023ydi}. This demonstrates that DNN can replace a GP even with only half the training data. Greater accuracy, however, still requires providing most of the data to the DNN for training.

In the case of 90\% training, 10\% testing, both models were able to reproduce their previous success. In fact, Scikit-Learn achieved a MSE for $v_2$ significantly smaller than before, $3.05 \times 10^{-6}$ down from $1.06 \times 10^{-5}$. However, $F_1$ demonstrates a possible issue with using this large training to testing ratio. The absolute MSE is up slightly, but more importantly the DNN error outlier for Scikit-Learn (TensorFlow) is $1.90 \times 10^{-1}$ ($9.53 \times 10^{-2}$). For TensorFlow that is a small increase from $6.30 \times 10^{-2}$, but for Scikit-Learn, the new outlier is more than five times its previous value of $2.64 \times 10^{-2}$. Thus, using a higher percentage of data for training can improve some results at the risk of having some extreme outliers.

This is by no means an exhaustive analysis of training to testing ratios. It demonstrates, however, that the frequently used 75/25 data splitting does indeed work well in our case.

\subsection{Caveats and future prospectives}\label{discuss}
A Gaussian Process typically provides two functions: accurate predictions of a simulation and an estimated error for each prediction, i.e. mean and variance. In this study, DNN were able to make accurate predictions. This is more impressive given the small dataset, and useful because of the long simulation times to create even a few data points. Unfortunately, the models used here were not capable of providing a variance for each prediction. The best method for accomplishing this is still an active topic of investigation, with possibilities ranging from measuring Monte Carlo dropout (MC-dropout) \cite{9010684, gal2016dropoutbayesianapproximationrepresenting} to deep ensembles \cite{lakshminarayanan2017simple} to mean-variance estimation neural networks \cite{374138, sluijterman2023optimal}. Future work should explore if any of these methods or others can be easily implemented without significantly increasing the training time, which currently takes only a few seconds for the models used here. For applications that do not require variance predictions, both Scikit-Learn and TensorFlow can be used to implement a quickly trainable DNN for emulations.

Above we have reported our best results with quantified variations, from the models that gave strong $R^2$ often, but some models behaved inconsistently. This is mainly due to our choice of randomly splitting the data (for a chosen training/testing ratio) and allowing the initial weights and biases of each model initialize randomly \emph{each time}. We chose this method because picking the optimal training/testing split would require prior knowledge that experiments cannot provide. So far as the DNN is concerned, the data \emph{is} random. Ideally, the DNN could be accurately trained regardless of which data is used for training and which for testing. We think that with a dataset this small, even slight differences in the training/testing split can lead to a very inconsistent DNN.  Oddly, using MAS to scale the data worsened this for the TensorFlow models despite decreasing the loss function, which is why we did not use it. This is a very frustrating setback after managing to avoid over-fitting, the most often cited problem for small datasets.

An additional challenge was the regular manual adjustments. In order to produce optimal results, we experimented with different data scalers, layer structures, activation functions, optimizers, number of iterations and epochs, and more. This manual adjustment allowed us to better learn and understand various features of DNN. For larger projects, however, it may be worth using an automated method such as another neural network that optimizes the hyperparameter setup for DNN as was done in \cite{Kasim_2022}, which used Scikit-Learn, or with the Keras Tuner \cite{omalley2019kerastuner} for TensorFlow.

\section{Summary}\label{sum}
To summarize, we found deep neural networks can accurately predict the values of the nuclear flow parameters $F_1$ and $v_2$ given the two input parameters X and K, even though only a small training dataset was provided. Moreover, it is about ten times more accurate than the GP emulator trained with the same dataset. 

The simplest of DNN models, however, is unable to fully replace a Gaussian emulator because it does not provide a variance prediction. The DNN used here was also incapable of solving directly the inverse problem to accurately infer X and K from $F_1$ and $v_2$ alone. On the other hand, the value of a faster and more reliable DNN emulator lies in its potential use in solving the inverse reaction problem within a Bayesian+DNN framework to infer the PDFs of model parameters. 

In the forward modeling with transport models, the whole phase space distribution functions of all particles in the final state of heavy-ion reactions are predicted. One can construct many observables to be compared with the experimental data in a Bayesian framework. To emulate multiple observables, one may need to generalize the simple DNN studied here to a committee of DNNs. Moreover, given a set of input parameters, transport model predictions for final observables are unique within statistical fluctuations. However, it is generally degenerate in the inverse process to go from a few final state observables to the input model parameters. Unraveling links between final state observables with input model parameters with inversion techniques has been a longstanding and challenging goal for the heavy-ion reaction community. Statistical inversions within the Bayesian framework using faster emulators for transport model simulators have been shown to be promising by the community. Our findings and experiences from constructing and testing a DNN as a faster and more reliable emulator of heavy-ion reactions at intermediate energies contribute positively to efforts in this direction.


\section*{Acknowledgements}
We would like to thank Jake Richter for allowing us to use his NN code to learn about neural networks and for giving us a reference point to see if ours worked. Many thanks also go to Plamen Krastev, Wen-Jie Xie and Kai Zhou for helpful discussions and communications. This research was sponsored in part by the U.S. Department of Energy under Award Number DE-SC0013702 and the TAMU-Commerce Division of Research and Economic Development.



\begin{thebibliography}{10}
\expandafter\ifx\csname url\endcsname\relax
  \def\url#1{\texttt{#1}}\fi
\expandafter\ifx\csname urlprefix\endcsname\relax\def\urlprefix{URL }\fi
\expandafter\ifx\csname href\endcsname\relax
  \def\href#1#2{#2} \def\path#1{#1}\fi

\bibitem{mcmc_ml}
C.~Andrieu, N.~de~Freitas, A.~Doucet, et~al., An introduction to mcmc for machine learning. {\JournalTitle{Machine Learning}}  {\bf50}, 5--43 (2003). \href {http://dx.doi.org/{https://doi.org/10.1023/A:1020281327116}} {doi:{https://doi.org/10.1023/A:1020281327116}}

\bibitem{il-st}
S.~Gazula, J.W. Clark, H.~Bohr, {Learning and prediction of nuclear stability by neural networks}. {\JournalTitle{Nucl. Phys. A}}  {\bf540}, 1--26 (1992). \href {http://dx.doi.org/10.1016/0375-9474(92)90191-L} {doi:10.1016/0375-9474(92)90191-L}

\bibitem{Boe22}
A.~Boehnlein, et~al., {Colloquium: Machine learning in nuclear physics}. {\JournalTitle{Rev. Mod. Phys.}}  {\bf94}, 031003 (2022). \href {http://arxiv.org/abs/2112.02309} {arXiv:2112.02309}, \href {http://dx.doi.org/10.1103/RevModPhys.94.031003} {doi:10.1103/RevModPhys.94.031003}

\bibitem{Zhou23}
K.~Zhou, L.~Wang, L.G. Pang, et~al., {Exploring QCD matter in extreme conditions with Machine Learning}. {\JournalTitle{Prog. Part. Nucl. Phys.}}  {\bf135}, 104084 (2024). \href {http://arxiv.org/abs/2303.15136} {arXiv:2303.15136}, \href {http://dx.doi.org/10.1016/j.ppnp.2023.104084} {doi:10.1016/j.ppnp.2023.104084}

\bibitem{He23}
W.~He, Q.~Li, Y.~Ma, et~al., {Machine learning in nuclear physics at low and intermediate energies}. {\JournalTitle{Sci. China Phys. Mech. Astron.}}  {\bf66}, 282001 (2023). \href {http://arxiv.org/abs/2301.06396} {arXiv:2301.06396}, \href {http://dx.doi.org/10.1007/s11433-023-2116-0} {doi:10.1007/s11433-023-2116-0}

\bibitem{Bass-nn}
S.A. Bass, A.~Bischoff, C.~Hartnack, et~al., {Neural networks for impact parameter determination}. {\JournalTitle{J. Phys. G}}  {\bf20}, L21--L26 (1994). \href {http://dx.doi.org/10.1088/0954-3899/20/1/004} {doi:10.1088/0954-3899/20/1/004}

\bibitem{WANG2022137508}
Y.~Wang, Z.~Gao, H.~Lu, et~al., {Decoding the nuclear symmetry energy event-by-event in heavy-ion collisions with machine learning}. {\JournalTitle{Physics Letters B}}  {\bf835}, 137508 (2022). \href {http://dx.doi.org/https://doi.org/10.1016/j.physletb.2022.137508} {doi:https://doi.org/10.1016/j.physletb.2022.137508}

\bibitem{UU}
Z.X. Yang, X.H. Fan, Z.P. Li, et~al., {A neural network approach for orienting heavy-ion collision events}. {\JournalTitle{Phys. Lett. B}}  {\bf848}, 138359 (2024). \href {http://arxiv.org/abs/2308.15796} {arXiv:2308.15796}, \href {http://dx.doi.org/10.1016/j.physletb.2023.138359} {doi:10.1016/j.physletb.2023.138359}

\bibitem{Rap20}
R.D. Lasseri, D.~Regnier, J.P. Ebran, et~al., Taming nuclear complexity with a committee of multilayer neural networks. {\JournalTitle{Phys. Rev. Lett.}}  {\bf124}, 162502 (2020). \href {http://dx.doi.org/10.1103/PhysRevLett.124.162502} {doi:10.1103/PhysRevLett.124.162502}

\bibitem{Pei21}
Z.A. Wang, J.~Pei, Optimizing multilayer bayesian neural networks for evaluation of fission yields. {\JournalTitle{Phys. Rev. C}}  {\bf104}, 064608 (2021). \href {http://dx.doi.org/10.1103/PhysRevC.104.064608} {doi:10.1103/PhysRevC.104.064608}

\bibitem{Dean22}
A.~Sarkar, D.~Lee, Self-learning emulators and eigenvector continuation. {\JournalTitle{Phys. Rev. Res.}}  {\bf4}, 023214 (2022). \href {http://dx.doi.org/10.1103/PhysRevResearch.4.023214} {doi:10.1103/PhysRevResearch.4.023214}

\bibitem{Love22}
A.E. Lovell, A.T. Mohan, T.M. Sprouse, et~al., Nuclear masses learned from a probabilistic neural network. {\JournalTitle{Phys. Rev. C}}  {\bf106}, 014305 (2022). \href {http://dx.doi.org/10.1103/PhysRevC.106.014305} {doi:10.1103/PhysRevC.106.014305}

\bibitem{Mol22}
O.M. Molchanov, K.D. Launey, A.~Mercenne, et~al., Machine learning approach to pattern recognition in nuclear dynamics from the ab initio symmetry-adapted no-core shell model. {\JournalTitle{Phys. Rev. C}}  {\bf105}, 034306 (2022). \href {http://dx.doi.org/10.1103/PhysRevC.105.034306} {doi:10.1103/PhysRevC.105.034306}

\bibitem{Kno23}
M.~Kn\"oll, T.~Wolfgruber, M.L. Agel, et~al., {Machine learning for the prediction of converged energies from ab initio nuclear structure calculations}. {\JournalTitle{Phys. Lett. B}}  {\bf839}, 137781 (2023). \href {http://arxiv.org/abs/2207.03828} {arXiv:2207.03828}, \href {http://dx.doi.org/10.1016/j.physletb.2023.137781} {doi:10.1016/j.physletb.2023.137781}

\bibitem{Nob23}
G.P.A. Nobre, D.A. Brown, S.J. Hollick, et~al., Novel machine-learning method for spin classification of neutron resonances. {\JournalTitle{Phys. Rev. C}}  {\bf107}, 034612 (2023). \href {http://dx.doi.org/10.1103/PhysRevC.107.034612} {doi:10.1103/PhysRevC.107.034612}

\bibitem{Yang23}
Y.L. Yang, P.W. Zhao, Deep-neural-network approach to solving the ab initio nuclear structure problem. {\JournalTitle{Phys. Rev. C}}  {\bf107}, 034320 (2023). \href {http://dx.doi.org/10.1103/PhysRevC.107.034320} {doi:10.1103/PhysRevC.107.034320}

\bibitem{Skyrme_DL}
N.~Hizawa, K.~Hagino, K.~Yoshida, {Analysis of a Skyrme energy density functional with deep learning}. {\JournalTitle{Phys. Rev. C}}  {\bf108}, 034311 (2023). \href {http://arxiv.org/abs/2306.11314} {arXiv:2306.11314}, \href {http://dx.doi.org/10.1103/PhysRevC.108.034311} {doi:10.1103/PhysRevC.108.034311}

\bibitem{Lay24}
D.~Lay, E.~Flynn, S.A. Giuliani, et~al., {Neural network emulation of spontaneous fission}. {\JournalTitle{Phys. Rev. C}}  {\bf109}, 044305 (2024). \href {http://arxiv.org/abs/2310.01608} {arXiv:2310.01608}, \href {http://dx.doi.org/10.1103/PhysRevC.109.044305} {doi:10.1103/PhysRevC.109.044305}

\bibitem{Kasim_2022}
M.F. Kasim, D.~Watson-Parris, L.~Deaconu, et~al., Building high accuracy emulators for scientific simulations with deep neural architecture search. {\JournalTitle{Machine Learning: Science and Technology}}  {\bf3}, 015013 (2021). \href {http://dx.doi.org/10.1088/2632-2153/ac3ffa} {doi:10.1088/2632-2153/ac3ffa}

\bibitem{Li-Bauer}
B.A. Li, C.M. Ko, W.~Bauer, {Isospin physics in heavy ion collisions at intermediate-energies}. {\JournalTitle{Int. J. Mod. Phys. E}}  {\bf7}, 147--230 (1998). \href {http://arxiv.org/abs/nucl-th/9707014} {arXiv:nucl-th/9707014}, \href {http://dx.doi.org/10.1142/S0218301398000087} {doi:10.1142/S0218301398000087}

\bibitem{LCK}
B.A. Li, L.W. Chen, C.M. Ko, {Recent Progress and New Challenges in Isospin Physics with Heavy-Ion Reactions}. {\JournalTitle{Phys. Rept.}}  {\bf464}, 113--281 (2008). \href {http://arxiv.org/abs/0804.3580} {arXiv:0804.3580}, \href {http://dx.doi.org/10.1016/j.physrep.2008.04.005} {doi:10.1016/j.physrep.2008.04.005}

\bibitem{pawel85}
P.~Danielewicz, G.~Odyniec, {Transverse Momentum Analysis of Collective Motion in Relativistic Nuclear Collisions}. {\JournalTitle{Phys. Lett. B}}  {\bf157}, 146--150 (1985). \href {http://arxiv.org/abs/2109.05308} {arXiv:2109.05308}, \href {http://dx.doi.org/10.1016/0370-2693(85)91535-7} {doi:10.1016/0370-2693(85)91535-7}

\bibitem{oll}
J.Y. Ollitrault, {Flow systematics from SIS to SPS energies}. {\JournalTitle{Nucl. Phys. A}}  {\bf638}, 195--206 (1998). \href {http://arxiv.org/abs/nucl-ex/9802005} {arXiv:nucl-ex/9802005}, \href {http://dx.doi.org/10.1016/S0375-9474(98)00413-8} {doi:10.1016/S0375-9474(98)00413-8}

\bibitem{art}
A.M. Poskanzer, S.A. Voloshin, Methods for analyzing anisotropic flow in relativistic nuclear collisions. {\JournalTitle{Phys. Rev. C}}  {\bf58}, 1671--1678 (1998). \href {http://dx.doi.org/10.1103/PhysRevC.58.1671} {doi:10.1103/PhysRevC.58.1671}

\bibitem{LRP}
A.~Sorensen, et~al., {Dense nuclear matter equation of state from heavy-ion collisions}. {\JournalTitle{Prog. Part. Nucl. Phys.}}  {\bf134}, 104080 (2024). \href {http://arxiv.org/abs/2301.13253} {arXiv:2301.13253}, \href {http://dx.doi.org/10.1016/j.ppnp.2023.104080} {doi:10.1016/j.ppnp.2023.104080}

\bibitem{Hermann}
H.~Wolter, et~al., {Transport model comparison studies of intermediate-energy heavy-ion collisions}. {\JournalTitle{Prog. Part. Nucl. Phys.}}  {\bf125}, 103962 (2022). \href {http://arxiv.org/abs/2202.06672} {arXiv:2202.06672}, \href {http://dx.doi.org/10.1016/j.ppnp.2022.103962} {doi:10.1016/j.ppnp.2022.103962}

\bibitem{GP}
C.~Rasmussen, C.~C.K. I.~Williams, {Gaussian Processes for Machine Learning}. {\JournalTitle{Cambridge, MA: MIT Press}} . \href {http://dx.doi.org/http://gaussianprocess.org/gpml/} {doi:http://gaussianprocess.org/gpml/}

\bibitem{LH}M.D. Mckay, R.J. Beckman, and W. J. Conover, {A Comparison of Three Methods for Selecting Values of Input Variables in the Analysis of Output from a Computer Code}.
 {\JournalTitle{Technometrics}} {\bf 42}, 55-61 (2000). \href{https://www.tandfonline.com/doi/abs/10.1080/00401706.2000.10485979}{https://doi.org/10.2307/1271432}

\bibitem{Li:2023ydi}
B.A. Li, W.J. Xie, {Bayesian inference of in-medium baryon-baryon scattering cross sections from HADES proton flow data}. {\JournalTitle{Nucl. Phys. A}}  {\bf1039}, 122726 (2023). \href {http://arxiv.org/abs/2303.10474} {arXiv:2303.10474}, \href {http://dx.doi.org/10.1016/j.nuclphysa.2023.122726} {doi:10.1016/j.nuclphysa.2023.122726}

\bibitem{HADES1}
J.~Adamczewski-Musch, et~al., {Directed, Elliptic, and Higher Order Flow Harmonics of Protons, Deuterons, and Tritons in $\mathrm{Au}+\mathrm{Au}$ Collisions at $\sqrt{{s}_{NN}}=2.4\text{ }\text{ }\mathrm{GeV}$}. {\JournalTitle{Phys. Rev. Lett.}}  {\bf125}, 262301 (2020). \href {http://arxiv.org/abs/2005.12217} {arXiv:2005.12217}, \href {http://dx.doi.org/10.1103/PhysRevLett.125.262301} {doi:10.1103/PhysRevLett.125.262301}

\bibitem{HADES2}
J.~Adamczewski-Musch, et~al., {Proton, deuteron and triton flow measurements in Au+Au collisions at $\sqrt{s_{_{{\text {NN}}}}}= 2.4$~GeV}. {\JournalTitle{Eur. Phys. J. A}}  {\bf59}, 80 (2023). \href {http://arxiv.org/abs/2208.02740} {arXiv:2208.02740}, \href {http://dx.doi.org/10.1140/epja/s10050-023-00936-6} {doi:10.1140/epja/s10050-023-00936-6}

\bibitem{Scott1}
J.~Novak, K.~Novak, S.~Pratt, et~al., {Determining Fundamental Properties of Matter Created in Ultrarelativistic Heavy-Ion Collisions}. {\JournalTitle{Phys. Rev. C}}  {\bf89}, 034917 (2014). \href {http://arxiv.org/abs/1303.5769} {arXiv:1303.5769}, \href {http://dx.doi.org/10.1103/PhysRevC.89.034917} {doi:10.1103/PhysRevC.89.034917}

\bibitem{Bass2}
J.E. Bernhard, P.W. Marcy, C.E. Coleman-Smith, et~al., {Quantifying properties of hot and dense QCD matter through systematic model-to-data comparison}. {\JournalTitle{Phys. Rev. C}}  {\bf91}, 054910 (2015). \href {http://arxiv.org/abs/1502.00339} {arXiv:1502.00339}, \href {http://dx.doi.org/10.1103/PhysRevC.91.054910} {doi:10.1103/PhysRevC.91.054910}

\bibitem{Scott2}
S.~Pratt, E.~Sangaline, P.~Sorensen, et~al., {Constraining the Eq. of State of Super-Hadronic Matter from Heavy-Ion Collisions}. {\JournalTitle{Phys. Rev. Lett.}}  {\bf114}, 202301 (2015). \href {http://arxiv.org/abs/1501.04042} {arXiv:1501.04042}, \href {http://dx.doi.org/10.1103/PhysRevLett.114.202301} {doi:10.1103/PhysRevLett.114.202301}

\bibitem{Ber16}
J.E. Bernhard, J.S. Moreland, S.A. Bass, et~al., {Applying Bayesian parameter estimation to relativistic heavy-ion collisions: simultaneous characterization of the initial state and quark-gluon plasma medium}. {\JournalTitle{Phys. Rev. C}}  {\bf94}, 024907 (2016). \href {http://arxiv.org/abs/1605.03954} {arXiv:1605.03954}, \href {http://dx.doi.org/10.1103/PhysRevC.94.024907} {doi:10.1103/PhysRevC.94.024907}

\bibitem{Scott3}
E.~Sangaline, S.~Pratt, {Toward a deeper understanding of how experiments constrain the underlying physics of heavy-ion collisions}. {\JournalTitle{Phys. Rev. C}}  {\bf93}, 024908 (2016). \href {http://arxiv.org/abs/1508.07017} {arXiv:1508.07017}, \href {http://dx.doi.org/10.1103/PhysRevC.93.024908} {doi:10.1103/PhysRevC.93.024908}

\bibitem{Bass1}
J.E. Bernhard, J.S. Moreland, S.A. Bass, {Bayesian estimation of the specific shear and bulk viscosity of quark\textendash{}gluon plasma}. {\JournalTitle{Nature Phys.}}  {\bf15}, 1113--1117 (2019). \href {http://dx.doi.org/10.1038/s41567-019-0611-8} {doi:10.1038/s41567-019-0611-8}

\bibitem{Weiss23}
B.~Weiss, J.F. Paquet, S.A. Bass, {Computational budget optimization for Bayesian parameter estimation in heavy-ion collisions}. {\JournalTitle{J. Phys. G}}  {\bf50}, 065104 (2023). \href {http://arxiv.org/abs/2301.08385} {arXiv:2301.08385}, \href {http://dx.doi.org/10.1088/1361-6471/acd0c7} {doi:10.1088/1361-6471/acd0c7}

\bibitem{Kuttan}M. O. Kuttan, J. Steinheimer, K. Zhou \textit{et al.}, {QCD Equation of State of Dense Nuclear Matter from a Bayesian Analysis of Heavy-Ion Collision Data},
\href{https://doi.org/10.1103/PhysRevLett.131.202303}{Phys. Rev. Lett. {\bf 131}, 202303 (2023)}.

\bibitem{Hef23}
M.R. Heffernan, C.~Gale, S.~Jeon, et~al., {Bayesian quantification of strongly interacting matter with color glass condensate initial conditions}. {\JournalTitle{Phys. Rev. C}}  {\bf109}, 065207 (2024). \href {http://arxiv.org/abs/2302.09478} {arXiv:2302.09478}, \href {http://dx.doi.org/10.1103/PhysRevC.109.065207} {doi:10.1103/PhysRevC.109.065207}

\bibitem{wang2024}
J.M. Wang, X.G. Deng, W.J. Xie, et~al., Bayesian inference of nuclear incompressibility from proton elliptic flow in central au+au collisions at 400 mev/nucleon. (2024).
\newblock \href {http://arxiv.org/abs/2406.07051} {arXiv:2406.07051}
\href{https://arxiv.org/abs/2406.07051}{https://arxiv.org/abs/2406.07051}

\bibitem{Phi21}
D.R. Phillips, et~al. (BAND Collaboration), {Get on the BAND Wagon: A Bayesian Framework for Quantifying Model Uncertainties in Nuclear Dynamics}. {\JournalTitle{J. Phys. G}}  {\bf48}, 072001 (2021). \href {http://arxiv.org/abs/2012.07704} {arXiv:2012.07704}, \href {http://dx.doi.org/10.1088/1361-6471/abf1df} {doi:10.1088/1361-6471/abf1df}

\bibitem{JETSCAPE}
W.~Fan, et~al. (JETSCAPE Collaboration), {New metric improving Bayesian calibration of a multistage approach studying hadron and inclusive jet suppression}. {\JournalTitle{Phys. Rev. C}}  {\bf109}, 064903 (2024). \href {http://arxiv.org/abs/2307.09641} {arXiv:2307.09641}, \href {http://dx.doi.org/10.1103/PhysRevC.109.064903} {doi:10.1103/PhysRevC.109.064903}

\bibitem{LiBA04A}
B.A. Li, C.B. Das, S.~Das~Gupta, et~al., {Momentum dependence of the symmetry potential and nuclear reactions induced by neutron rich nuclei at RIA}. {\JournalTitle{Phys. Rev. C}}  {\bf69}, 011603 (2004). \href {http://arxiv.org/abs/nucl-th/0312032} {arXiv:nucl-th/0312032}, \href {http://dx.doi.org/10.1103/PhysRevC.69.011603} {doi:10.1103/PhysRevC.69.011603}

\bibitem{LiBA04B}
B.A. Li, C.B. Das, S.~Das~Gupta, et~al., {Effects of momentum dependent symmetry potential on heavy ion collisions induced by neutron rich nuclei}. {\JournalTitle{Nucl. Phys. A}}  {\bf735}, 563--584 (2004). \href {http://arxiv.org/abs/nucl-th/0312054} {arXiv:nucl-th/0312054}, \href {http://dx.doi.org/10.1016/j.nuclphysa.2004.02.016} {doi:10.1016/j.nuclphysa.2004.02.016}

\bibitem{Li-Chen}
B.A. Li, L.W. Chen, {Nucleon-nucleon cross sections in neutron-rich matter and isospin transport in heavy-ion reactions at intermediate energies}. {\JournalTitle{Phys. Rev. C}}  {\bf72}, 064611 (2005). \href {http://arxiv.org/abs/nucl-th/0508024} {arXiv:nucl-th/0508024}, \href {http://dx.doi.org/10.1103/PhysRevC.72.064611} {doi:10.1103/PhysRevC.72.064611}

\bibitem{scikit-learn}
F.~Pedregosa, G.~Varoquaux, A.~Gramfort, et~al., Scikit-learn: Machine learning in {P}ython. {\JournalTitle{Journal of Machine Learning Research}}  {\bf12}, 2825--2830 (2011).

\bibitem{sklearn_api}
L.~Buitinck, G.~Louppe, M.~Blondel, et~al., in \emph{ECML PKDD Workshop: Languages for Data Mining and Machine Learning}, {API} design for machine learning software: experiences from the scikit-learn project. 2013, pp. 108--122

\bibitem{tensorflow2015-whitepaper}
M.~Abadi, A.~Agarwal, P.~Barham, et~al., {TensorFlow}: Large-scale machine learning on heterogeneous systems., software available from tensorflow.org (2015)
\href{https://www.tensorflow.org/}{https://www.tensorflow.org/}

\bibitem{chollet2015keras}
F.~Chollet, Keras., \url{https://github.com/fchollet/keras} (2015)

\bibitem{Liu:1989esw}
D.C. Liu, J.~Nocedal, {On the limited memory BFGS method for large scale optimization}. {\JournalTitle{Math. Programming}}  {\bf45}, 503--528 (1989). \href {http://dx.doi.org/10.1007/BF01589116} {doi:10.1007/BF01589116}

\bibitem{kingma2017adammethodstochasticoptimization}
D.P. Kingma, J.~Ba, Adam: A method for stochastic optimization. (2017).
\newblock \href {http://arxiv.org/abs/1412.6980} {arXiv:1412.6980}
\href{https://arxiv.org/abs/1412.6980}{https://arxiv.org/abs/1412.6980}

\bibitem{Richter:2023zec}
J.~Richter, B.A. Li, {Empirical radius formulas for canonical neutron stars from bidirectionally selecting features of equations~of state in extended Bayesian analyses of observational data}. {\JournalTitle{Phys. Rev. C}}  {\bf108}, 055803 (2023). \href {http://arxiv.org/abs/2307.05848} {arXiv:2307.05848}, \href {http://dx.doi.org/10.1103/PhysRevC.108.055803} {doi:10.1103/PhysRevC.108.055803}

\bibitem{Ber88}
G.~F.~Bertsch, G.~E.~Brown, V.~Koch and B.~A.~Li,
Pion Collectivity in Relativistic Heavy Ion Collisions,
\href {https://doi.org/10.1016/0375-9474(88)90024-3}{Nucl. Phys. A \textbf{490}, 745 (1988).} 

\bibitem{Zheng99}
Y.~M.~Zheng, C.~M.~Ko, B.~A.~Li and B.~Zhang,
Elliptic flow in heavy ion collisions near the balance energy,
\href{https://journals.aps.org/prl/pdf/10.1103/PhysRevLett.83.2534}{Phys. Rev. Lett. \textbf{83},2534 (1999).}

\bibitem{LiSustich}
B.~A.~Li and A.~T.~Sustich,
Differential flow in heavy ion collisions at balance energies,
\href{https://journals.aps.org/prl/pdf/10.1103/PhysRevLett.82.5004}{Phys. Rev. Lett. \textbf{82}, 5004 (1999).} 

\bibitem{Dan02}
P.~Danielewicz,
Hadronic transport models,
\href{https://inspirehep.net/files/2781c2e8352c7e876504e261ce6afbde}{Acta Phys. Polon. B \textbf{33}, 45 (2002).}

\bibitem{BALI2}
B.~A.~Li, B.~J.~Cai, L.~W.~Chen and J.~Xu,
Nucleon Effective Masses in Neutron-Rich Matter,
\href{https://www.sciencedirect.com/science/article/pii/S0146641018300012?via%3Dihub}{Prog. Part. Nucl. Phys. \textbf{99}, 29 (2018).}

\bibitem{Herman1}T.~Gaitanos, C.~Fuchs and H.~H.~Wolter,
Nuclear stopping and flow in heavy ion collisions and the in-medium NN cross section,
\href{https://www.sciencedirect.com/science/article/abs/pii/S0370269305001486?via%3Dihub}
{Phys. Lett. B \textbf{609}, 241 (2005).} 

\bibitem{Zhang07}
Y.~Zhang, Z.~Li and P.~Danielewicz,
In-medium NN cross-sections determined from stopping and collective flow in intermediate-energy heavy-ion collisions,
\href{https://journals.aps.org/prc/abstract/10.1103/PhysRevC.75.034615}{Phys. Rev. C \textbf{75}, 034615 (2007).}

\bibitem{PLi18}
P.~Li, Y.~Wang, Q.~Li, C.~Guo and H.~Zhang,
Effects of the in-medium nucleon-nucleon cross section on collective flow and nuclear stopping in heavy-ion collisions in the Fermi-energy domain, \href{https://journals.aps.org/prc/pdf/10.1103/PhysRevC.97.044620}{Phys. Rev. C \textbf{97}, 044620 (2008).}

\bibitem{Li22}
P.~Li, Y.~Wang, Q.~Li and H.~Zhang,
Accessing the in-medium effects on nucleon-nucleon elastic cross section with collective flows and nuclear stopping,
\href{https://www.sciencedirect.com/science/article/pii/S0370269322001538?via%3Dihub}{Phys. Lett. B \textbf{828}, 137019 (2022).}

\bibitem{Gale}D. Persram and C. Gale, Elliptic flow in intermediate-energy heavy ion collisions and in-medium effects,
\href{https://journals.aps.org/prc/abstract/10.1103/PhysRevC.65.064611}{
Phys. Rev. C{\bf 65}, 064611 (2002).}

\bibitem{Fuchs01}
C.~Fuchs, A.~Faessler and M.~El-Shabshiry,
Off-shell behavior of the in-medium nucleon-nucleon cross-section,
\href{https://journals.aps.org/prc/abstract/10.1103/PhysRevC.64.024003}{Phys. Rev. C \textbf{64}, 024003 (2001).}

\bibitem{Joseph_2022}
J.V.~Roshan, Optimal ratio for data splitting, \href{https://doi.org/10.1002/sam.11583}{Stat. Anal. Data Min.: ASA Data Sci. J. {\bf 15} (2022).}

\bibitem{9010684}
J.~Postels, F.~Ferroni, H.~Coskun, et~al., in \emph{2019 IEEE/CVF International Conference on Computer Vision (ICCV)}, Sampling-free epistemic uncertainty estimation using approximated variance propagation. 2019, pp. 2931--2940.
\newblock \href {http://dx.doi.org/10.1109/ICCV.2019.00302} {doi:10.1109/ICCV.2019.00302}

\bibitem{gal2016dropoutbayesianapproximationrepresenting}
Y.~Gal, Z.~Ghahramani, Dropout as a bayesian approximation: Representing model uncertainty in deep learning. (2016).
\newblock \href {http://arxiv.org/abs/1506.02142} {arXiv:1506.02142}
\href{https://arxiv.org/abs/1506.02142}{https://arxiv.org/abs/1506.02142}

\bibitem{lakshminarayanan2017simple}
B.~Lakshminarayanan, A.~Pritzel, C.~Blundell, Simple and scalable predictive uncertainty estimation using deep ensembles. (2017).
\newblock \href {http://arxiv.org/abs/1612.01474} {arXiv:1612.01474}

\bibitem{374138}
D.~Nix, A.~Weigend, in \emph{Proceedings of 1994 IEEE International Conference on Neural Networks (ICNN'94)}, Estimating the mean and variance of the target probability distribution. Vol.~1, 1994, pp. 55--60 vol.1.
\newblock \href {http://dx.doi.org/10.1109/ICNN.1994.374138} {doi:10.1109/ICNN.1994.374138}

\bibitem{sluijterman2023optimal}
L.~Sluijterman, E.~Cator, T.~Heskes, Optimal training of mean variance estimation neural networks. (2023).
\newblock \href {http://arxiv.org/abs/2302.08875} {arXiv:2302.08875}

\bibitem{omalley2019kerastuner}
T.~O'Malley, E.~Bursztein, J.~Long, et~al., Kerastuner., \url{https://github.com/keras-team/keras-tuner} (2019)

\end{thebibliography}
\end{document}